\documentclass[12pt]{article} 

\usepackage{amsmath,amssymb,amsfonts}
\usepackage{url}
\usepackage{graphicx}
\usepackage{tikz}
\usepackage{cite}
\usepackage{multicol}

\newcommand{\mean}[1]{\left\langle #1 \right\rangle}

\newcommand{\pr}{^\prime}
\newcommand{\prsq}{^{\prime\:2}}
\newcommand{\ns}{^\mathrm{(ns)}}
\newcommand{\sca}{^\mathrm{(sc)}}
\newcommand{\prsca}{^{\prime\:\mathrm{(sc)}}}

\begin{document}

\title{The maximum penalty criterion for ridge regression: application to the calibration of the force constant in elastic network models.}

\author{
  Ugo Bastolla$^{(1)}$ and Yves Dehouck$^{(2)}$\\
  \small $^{(1)}$ Centro de Biologia Molecular "Severo Ochoa"\\
  \small CSIC-UAM Cantoblanco, 28049 Madrid, Spain. E-mail: ubastolla@cbm.csic.es \vspace{0.1cm}\\
  \small $^{(2)}$ Machine Learning Group, Universit\'e Libre de Bruxelles (ULB).\\
  \small Boulevard du Triomphe CP 212, 1050 Brussels, Belgium. E-mail: ydehouck@ulb.ac.be
}

\date{}
\maketitle

\begin{abstract}
Multivariate regression is a widespread computational technique that may give meaningless results if the explanatory variables are too numerous or highly collinear.
Tikhonov regularization, or ridge regression, is a popular approach to address this issue. 
We reveal here a formal analogy between ridge regression and statistical mechanics, where the objective function is comparable to a free energy, and the ridge parameter plays the role of temperature.
This analogy suggests two new criteria to select a suitable ridge parameter: the specific-heat (Cv) and the maximum penalty (MP) fits. 
We apply these methods to the calibration of the force constant in elastic network models (ENM).
This key parameter, which determines the amplitude of the predicted atomic fluctuations, is commonly obtained by fitting crystallographic B-factors.
However, rigid-body motions are typically partially neglected in such fits, even though their importance has been repeatedly stressed.
Considering the full set of rigid-body and internal degrees of freedom bears significant risks of overfitting, due to the strong correlations between explanatory variables, and requires thus careful regularization.
Using simulated data, we show that ridge regression with the Cv or MP criterion markedly reduces the error of the estimated force constant, its across-protein variation, and the number of proteins with unphysical values of the fit parameters, in comparison with popular regularization schemes such as generalized cross-validation.
When applied to protein crystals, the new methods are shown to provide a more robust calibration of ENM force constants, even though our results indicate that rigid-body motions account on average for more than 80\% of the amplitude of B-factors.
While MP emerges as the optimal choice for fitting crystallographic B-factors, the Cv fit is more robust to the nature of the data, and is thus an interesting candidate for other applications.
\end{abstract}

\clearpage

There is growing interest in the investigation of the intrinsic dynamical properties of proteins in their native state,
for these properties play a key role in ensuring proper functional activity, 
notably for catalysis \cite{Kern2005}, allosteric regulation \cite{Goodey_Benkovic_2008,allostery}, or molecular recognition \cite{conf_selection}.
However, despite recent progress \cite{Kern2010}, the experimental study of protein dynamics remains rather challenging,
and computational methods can thus often provide valuable alternatives.
Among those, elastic network models (ENM) \cite{Tirion96,Atilgan01,Bahar_review_05} are becoming increasingly popular,
since they are able to provide detailed analytic predictions of native protein dynamics at a very reasonable computational cost.
The ENM predictions have been shown to correlate well with experimentally observed conformational changes \cite{Tama2001,binding,Bastolla2013},
and with long molecular dynamics trajectories \cite{Orozco07,CHARMMvsENM}.

One of the advantages of ENMs is that their force field is derived from the experimentally determined structure of the protein of interest,
adopting the principle of minimal frustration, and relies on a very small number of coarse-grained parameters.
Typically, pairs of residues separated by a spatial distance lower than a certain cut-off value are 
identified as relevant interactions and connected by elastic springs.
The dynamical behaviour of the resulting network is determined by the stiffness of the spring assigned to each pair of residues,
which is often expressed as a constant multiplied by a decaying power function of the interresidue distance.
Several previous studies have been focused on the determination of the optimal parameters of this model,
i.e. the distance-dependence of the force constant, via the cut-off and the exponential factor \cite{Hinsen00,Moritsugu07,Yang09,Dehouck13},
or the influence of the chemical nature of each amino acid type \cite{Dehouck13,Hamacher06,Gerek09}.
In the present work, we address a somewhat different question, the evaluation of the overall scale of the force constants.
Although it does not influence the shape of the normal modes,
this parameter is crucial for determining the amplitude of the predicted internal motions of the macromolecule.
For that purpose, we adopt here the torsional network model (TNM), an ENM in torsion angle space that preserves the bond lengths and bond angles within the protein \cite{TNM}.

In ENM studies, it is customary to obtain the scale of the force constants by fitting the predicted thermal displacements of each atom to
the experimental mean square fluctuations $\mean{\Delta r_i^2}$ measured as temperature factors (B-factors) in X-ray crystallography.
This approach is based on the implicit assumption that the atomic displacements underlying crystallographic B-factors
result mainly from motions due to internal degrees of freedom.
However, it has been known for a long time that the B-factors are mainly influenced by rigid-body motions taking place in the crystal \cite{Kuriyan1991}.
On the other hand, crystallised macromolecules experience a different environment than when isolated in solution,
and the contacts established with neighbouring molecules in the crystal have also been shown to affect the normal modes of motion \cite{Hinsen2008,Riccardi2009}.
Several studies, in which crystal contacts were modelled explicitly, did however come to the conclusion that these contacts only weakly perturbate the internal dynamics of the protein,
while the anisotropic temperature factors are dominated by rigid-body motions of the protein \cite{Soheilifard2008,Hafner2010,Lezon2012}.
Soheilifard et al. \cite{Soheilifard2008}, and later Lezon \cite{Lezon2012}, proposed to improve the fit of B-factors
by considering the amplitudes of rigid-body motions via six additional fitting parameters.
However, the proposed fits are not complete, because ten parameters are necessary for a full representation of
the thermal fluctuations due to rigid-body motions \cite{TLS}.

It is in principle straightforward to perform a complete fit of B-factors using 10 free parameters corresponding to rigid-body motions, plus one free parameter that rescales the internal motions predicted by the ENM, which is equal to the inverse of the force constant.
In general, there is however a high level of collinearity between the variables describing internal motions and rigid-body rotations.
A fit of B-factors that fully accounts for rigid-body motions runs therefore a significant risk of overfitting and must be carefully regularized.
Ridge regression \cite{ridge} is one of the most common methods for regularizing fits with many variables.
It relies heavily on the choice of an adequate value for the ridge parameter but, although several criteria have been proposed for that purpose \cite{GCV,RR,L-curve}, 
there is no consensus on how to systematically determine the optimal value of this parameter.
In this paper, we propose two new criteria for choosing the ridge parameter, based on the analogy between ridge regression and statistical mechanics.
We call maximum penalty (MP) or specific-heat (Cv) fit the ridge regression performed with either choice of the ridge parameter.
We show that the MP fit yields close to optimal results when rigid-body motions account for a fraction of the fluctuations close to that estimated for X-ray structures,
while the Cv fit is more robust to the amplitude of rigid-body fluctuations.
In contrast, other widely used approaches, such as the generalized cross-validation (GCV) criterion \cite{GCV}, fail to provide a sufficient level of regularization.
The programs for performing the MP and Cv fits and computing the force constants within the ENM are available on request.

\section*{Analytical results}

\subsection*{Complete fit of B-factors}
Before addressing the regularization of the B-factor fit, we present the equations that describe it.
We assume that the thermal displacements are the combination of internal plus rigid-body motions,
$\Delta \vec{r}_i=\Delta \vec{r}_i^{\:\mathrm{int}}+\vec{\omega}\times \vec{r}_i+\vec{t}$.
Independence between rigid-body motions and internal motions results in the thermal averages
$\mean{\left|\Delta \vec{r}_i\right|^2}=
\mean{\left|\Delta \vec{r}_i\right|^2}_{\mathrm{int}}+
\mean{\left|\vec{t}\right|^2}+r_i\times \mean{2\vec{\omega}\times \vec{t}}+
\left(\vec{r}_i,\mathbf{I}\vec{r}_i\right)$, which lead to the fit
\begin{equation}
\mean{\left|\Delta \vec{r}_i\right|^2}\approx
a\pr_{\mathrm{trasl}} + \sum_{x=1,3}a\pr_{1x}r_{ix}+\sum_{x<=y}a\pr_{2xy}r_{ix}r_{iy} +
a\pr_{\mathrm{ENM}}\mean{\left|\Delta \vec{r}_i\right|^2}_{\mathrm{ENM}}
\label{eq:A}
\end{equation}

\noindent
We rename the parameters as $a\pr_k$ and rewrite the fit as
$y\pr_i = \mean{\left|\Delta \vec{r}_i\right|^2}_{\mathrm{Bfact}} \approx \sum_{k=0}^{10} X\pr_{ik}a\pr_k$,
with $X\pr_{i0}=1$, $X\pr_{ik}=r_{ik}$ for $k=1,2,3$, and $X\pr_{ik}=r_{ix}r_{iy}$ for $k=4\cdots 9$ ($x$ and $y=1\cdots 3$, with $x\leq y$).
Here and in the following, we denote by $i=1\cdots N$ any of the $N$ atoms and $k=0\cdots 10$ any of the $P=11$ variables.
The variable $X\pr_{i10}=\mean{\left|\Delta \vec{r}_i\right|^2}_{\mathrm{ENM}}$ corresponds to the thermal fluctuations due to internal motions, predicted by the ENM (see Methods).
The force constant of the model is thus obtained directly from the fit, as $\kappa=1/a\pr_{\mathrm{ENM}}$.

In the context of B-factors, it is natural to weight, by its mass $m_i$, the contribution of each atom $i$ to the error of the fit.
We use therefore the weighted variables $\sqrt{m_i}\:X\pr_{ik}$ and $y_i=\sqrt{m_i}\:y\pr_i$ instead of $X\pr_{ik}$ and $y\pr_i$, respectively
(this does not make any difference if, as it is customary, only the alpha carbons are considered).
Moreover, it is convenient to express the fit in terms of the normalized and dimensionless variables $X_{ik}=\sqrt{m_i}\:X\pr_{ik}/\sqrt{\sum_j m_j X\prsq_{jk}}$
and the normalized and dimensionless parameters $a_k=a\pr_k \sqrt{\sum_j m_j X\prsq_{jk}}$.
In matrix notations, the fit is then simply written $\mathbf{y} \approx \mathbf{Xa}$.

\subsection*{Ridge regression}

To limit the risk of overfitting, we adopt the Tychonov regularization, also known as ridge regression.
We first describe the usual approach for obtaining the non-scaled parameters $a\ns_k$,
which are defined as the values of the fit parameters $a_k$ that minimize the quantity
\begin{eqnarray}
G\ns(\Lambda, \mathbf{X}, \mathbf{y}, \mathbf{a}) & =& 
\left((\mathbf{Xa}-\mathbf{y}),(\mathbf{Xa}-\mathbf{y})\right)+\Lambda\left(\mathbf{a},\mathbf{a}\right) \\
& = & \left(\mathbf{a}, \mathbf{Ca}\right) - 2\left(\mathbf{a},\mathbf{X}^T \mathbf{y}\right)
+\left(\mathbf{y},\mathbf{y}\right) +\Lambda\left(\mathbf{a},\mathbf{a}\right)\, ,\nonumber
\end{eqnarray}
where $\left(\mathbf{x},\mathbf{y}\right)$ denotes the scalar product and $\mathbf{C}=(\mathbf{X}^T \mathbf{X})$ is the covariance matrix.
The ordinary least square (OLS) regression is recovered as the special case $\Lambda=0$.
The minimization of $G\ns$ can be interpreted as a constrained minimization of the error of the fit $E=\left((\mathbf{Xa}-\mathbf{y}),(\mathbf{Xa}-\mathbf{y})\right)$,
where the constraint is set on the norm of the parameters, $\left(\mathbf{a},\mathbf{a}\right)$, via a Lagrange multiplier $\Lambda$.
In other words, the error cannot be minimized at the cost of having too large values of the parameters.
The explicit solution of the above minimization problem is $\mathbf{a}\ns=\left(\mathbf{C}+\Lambda \mathbf{I}\right)^{-1}\left(\mathbf{X}^T \mathbf{y}\right)$.

Since the covariance matrix $\mathbf{C}$ is symmetric and positive definite, its eigenvalues $\lambda_\alpha$ are real and positive.
To simplify the computation, we define the normalized projections of the fitted variable $\mathbf{y}$ over the eigenvectors $\mathbf{u}^\alpha$ of the covariance matrix as $y^\alpha=\left(\mathbf{X}^T \mathbf{y}, \mathbf{u}^\alpha\right)$,
i.e. $y^\alpha=\sum_{k}\left(\sum_i X_{ik} y_i\right)u^\alpha_k$.
The solution is then given by the following formula,
which is convenient if we have to perform computations for several values of the Tykhonov parameter $\Lambda$:
\begin{equation}
a\ns_k = \sum_\alpha \frac{y^\alpha u^\alpha_k}{\lambda_\alpha+\Lambda}\, .
\label{eq:A_spectral}
\end{equation}
When $\Lambda$ increases, the parameters $a\ns_k$ tend to zero and so does the fitted dependent variable.
Protocols for ridge regression typically
address this problem by avoiding to penalize the offset of the fit, and choosing this offset in such a way that the fit is unbiased,
i.e. that the average of the fit is equal to the average of $\mathbf{y}$.
Nevertheless, this procedure modifies the relationship between the explanatory variables.
In particular, in the present case, the offset has to be interpreted as the component of the fit due to translations.
Increasing the offset would have the effect of artificially increasing the contribution of translations,
which would then be treated differently from the other degrees of freedom.

\subsection*{Rescaled ridge regression}
To ensure that the fitted dependent variable is correctly scaled with respect to $\mathbf{y}$,
while still considering translational motions similarly to other degrees of freedom,
we modify the ridge regression protocol so as to optimize the scale of the fit parameters $a_k$.
More precisely, we multiply all parameters $a\ns_k$ by a constant scalar $\nu$, to obtain the rescaled parameters $a\sca_k$.
This transformation does not modify the physical interpretation of the fit.
It is easy to see that the optimally rescaled parameters have to satisfy the condition $H(\mathbf{X},\mathbf{y},\mathbf{a}\sca)=0$, with 
\begin{equation}
H( \mathbf{X}, \mathbf{y}, \mathbf{a})\equiv \left(\mathbf{X}^T \mathbf{y},\mathbf{a}\right)-\left(\mathbf{a}, \mathbf{Ca}\right)\, .
\label{eq:scale}
\end{equation}
In order to keep the analytic treatibility, we impose this constraint on the scale via a new Lagrange multiplier $\mu$,
so that the rescaled ridge regression is still formulated as the minimization of a quadratic function of the parameters:

\begin{eqnarray}
G\prsca(\Lambda, \mu, \mathbf{X}, \mathbf{y}, \mathbf{a}) & =& 
E + (1-\mu)\Lambda \left(\mathbf{a,a}\right) +\mu H \label{eq:scaledpr}\\
& = & (1-\mu)\left[ \left(\mathbf{a}, \mathbf{Ca}\right)-2\nu\left(\mathbf{a},\mathbf{X}^T \mathbf{y}\right)
+ \Lambda\left(\mathbf{a},\mathbf{a}\right)\right]
+ \left(\mathbf{y},\mathbf{y}\right)\, , \nonumber
\end{eqnarray}
with $\nu=(1-\mu/2)/(1-\mu)$.
The non-scaled fit can be obtained as a particular case by setting $\nu(\Lambda)\equiv 1$, which implies $\mu=0$.
Since the term $(\mathbf{y},\mathbf{y})$ is constant,
minimizing the objective function $G\prsca$ is equivalent to minimizing $G\prsca/(1-\mu)$,
and the solution of this problem is thus given by $\mathbf{a}\sca=\nu\left(\mathbf{C}+\Lambda \mathbf{I}\right)^{-1}\left(\mathbf{X}^T \mathbf{y}\right)$, that is
\begin{equation}
a\sca_k \equiv \nu(\Lambda) a\ns_k
=\nu(\Lambda) \sum_\alpha \frac{y^\alpha u^\alpha_k}{\lambda_\alpha+\Lambda}\, .
\label{eq:Asca}
\end{equation}
The proportionality factor $\nu$ must fulfill $H=0$ (Eq.(\ref{eq:scale})). We find
\begin{eqnarray}
& & \nu(\Lambda)=\frac{\left(\mathbf{X}^T \mathbf{Y},\mathbf{a}\ns\right)}{\left(\mathbf{a}\ns, \mathbf{C}\mathbf{a}\ns\right)}=
1+\Lambda \eta(\Lambda) \\
& & \eta(\Lambda)=
\frac{\sum_\alpha \frac{(y^\alpha)^2}{(\lambda_\alpha+\Lambda)^2}}
{\sum_\alpha \lambda_\alpha\frac{(y^\alpha)^2}{(\lambda_\alpha+\Lambda)^2}}\, .
\label{eq:nu}
\end{eqnarray}
Note that we defined the penalisation term as $(1-\mu)\Lambda \left(\mathbf{a,a}\right)$ instead of $\Lambda \left(\mathbf{a,a}\right)$.
This adjustment is of course not necessary, but it is convenient as it allows the rescaled solution to be proportional to the non-scaled solution obtained at the same value of $\Lambda$.

The interplay of the constraints imposed by the two Lagrange multipliers implies that,
contrary to $a\ns_k$, the rescaled fit parameters $a\sca_k$ do not tend to zero when $\Lambda$ increases.
\begin{eqnarray}
a\sca_{k,\infty} \equiv \lim_{\Lambda\rightarrow \infty} a\sca_k & = &
\eta_{\infty} \sum_\alpha y^\alpha u^\alpha_k
\, ,\, \textrm{with} \label{eq:ascinf} \\
\eta_{\infty} \equiv \lim_{\Lambda\rightarrow \infty} \eta(\Lambda) & = & \frac{\sum_\alpha (y^\alpha)^2}{\sum_\alpha (y^\alpha)^2\lambda_\alpha}\, . \label{eq:nuinf}
\end{eqnarray}
These limit values of the parameters $a\sca_k$
are independent of the correlation matrix $\mathbf{C}$,
except for their scale $\eta_{\infty}$ that depends on the eigenvalues $\lambda_\alpha$.
Therefore, when $\Lambda$ increases, the information on the correlations between the predictor variables is progressively lost,
and their relative weights in the fit become more and more strongly determined by their individual correlations with the dependent variable.

There are alternative ways to formulate the rescaled ridge regression problem.
In particular, instead of the constraint on the norm of the parameters $(\mathbf{a},\mathbf{a})$, we may choose to impose a constraint
on the Euclidian distance $F$ between the parameters $\mathbf{a}$
and adequately chosen reference values of these parameters, $\mathbf{a}^\circ$:
\begin{equation}
F=\left((\mathbf{a}-\mathbf{a}^\circ),(\mathbf{a}-\mathbf{a}^\circ)\right)\, . \label{eq:F}
\end{equation} 
Provided that the reference parameters are chosen as $\mathbf{a}^\circ=\xi(\Lambda) \mathbf{X}^T \mathbf{y}=\xi(\Lambda)  \sum_\alpha y^\alpha u^\alpha$,
the minimization with respect to $\mathbf{a}$ of the new objective function 
\begin{eqnarray}
G\sca(\Lambda, \nu, \mathbf{X}, \mathbf{y}, \mathbf{a}) & =& 
E + (1-\mu)\Lambda F +\mu H \label{eq:scaled}\\
& = & (1-\mu)\left[ \left(\mathbf{a}, \mathbf{Ca}\right)-2\nu\left(\mathbf{a},\mathbf{X}^T \mathbf{y}\right)
+ \Lambda\left(\mathbf{a},\mathbf{a}\right) + \Lambda\left(\mathbf{a}^\circ,\mathbf{a}^\circ\right)\right]
+ \left(\mathbf{y},\mathbf{y}\right)\, , \nonumber
\end{eqnarray}
yields $\mathbf{a}\sca=\nu\left(\mathbf{C}+\Lambda \mathbf{I}\right)^{-1}\left(\mathbf{X}^T \mathbf{y}\right)$,
which is exactly the same result as Eq.(\ref{eq:Asca}) above.
Note that the previous formulation of the objective function, in Eq.(\ref{eq:scaledpr}), is retrieved by setting $\xi = \xi_0 \equiv 0$.
The parameter that enforces the scale is now
\begin{equation}
\nu(\Lambda)=\Lambda\xi + (1-\mu/2)/(1-\mu)= 1+\Lambda\eta(\Lambda)
\label{eq:mu}
\end{equation}
Note that the value of the scale parameter is still determined by Eq.(\ref{eq:nu}), but we are free to choose the multiplier $\mu(\Lambda)$ and the scale of the reference parameters $\xi(\Lambda)$ as it is most convenient. 
A simple and interesting choice is to set $\xi(\Lambda)$ independent of $\Lambda$. To recover the correct infinite $\Lambda$ limit, it must hold $\xi=\xi_{\infty} \equiv \eta_{\infty}$,
so that $\mathbf{a}^\circ = \mathbf{a}\sca_{\infty}$, and the term $F$ vanishes in the limit of infinite $\Lambda$.
This gives a more straightforward interpretation to the penalisation term in rescaled ridge regression,
i.e. the error cannot be minimized at the cost of having parameters values too different from those that would be obtained if
the predictor variables were not correlated to each other.
Another possibility is to set $\mu$ as constant. Again, the infinite $\Lambda$ limit requires $\mu=0$ and $\xi(\Lambda)=\xi_{\Lambda} \equiv (\nu(\Lambda)-1)/\Lambda$. Since $\mu=0$, we do not have to impose any constraint on the scale, but the optimal scale is automatically imposed by the chosen scale of the reference paraters.
In that case, the reference parameters are not constant but depend on $\Lambda$ via $\xi_{\Lambda}$.
There is also a direct correspondence with the non-scaled ridge regression since, in that case,
both $\xi\ns_{\infty}=0$ and $\xi\ns_{\Lambda}=0$, and $\mathbf{a}^{\circ\:\mathrm{(ns)}} \equiv 0$.

\subsection*{Statistical mechanics analogy}

In view of addressing the problem of the optimal choice of $\Lambda$,
we note that there is a formal analogy between the function $G$ that has to be minimized and a free energy,
where the ridge parameter $\Lambda$ plays the role of the temperature. 

Consider a discrete physical system that can access $K$ microstates.
The Boltzmann distribution in statistical mechanics is given by $p_k=\exp(-E_k/k_BT)/\sum_j \exp(-E_j/k_BT)$,
where $p_k$ is the probability of microstate $k$ and $E_k$ its energy, $k_B$ is the Boltzmann constant, and $T$ the absolute temperature.
It can be formally described as the probability distribution that maximizes the entropy for given average energy or,
equivalently, minimizes the energy for given entropy.
The constraint on the entropy is fixed by the Lagrange multiplier $T$,
and the normalization condition on the probabilities, $\sum_k p_k=1$, is imposed through another Lagrange multiplier.
The objective function can be put in the form $G(\mathbf{p})=E(\mathbf{p})-TS(\mathbf{p})$,
where $E(\mathbf{p})=\sum_k p_k E_k$ is the average energy and $S(\mathbf{p})=-k_B \sum_k p_k\log(p_k)$ is the entropy.
For the sake of the analogy, we adopt the equivalent objective function $G(\mathbf{p})=E(\mathbf{p})+T(S(\mathbf{p}^\circ)-S(\mathbf{p}))$,
where $p^\circ_k=1/K$ is the uniform distribution in the space of the microstates.
The term $S(\mathbf{p}^\circ)$ does not modify the result of the constrained minimization.
The term $S(\mathbf{p}^\circ)-S(\mathbf{p})$ is proportional to the Kullback-Leibler divergence ($D_{\mathrm{KL}}$) between the Boltzmann distribution $p_k(T)$
and the reference distribution $p^\circ_k=1/K$ that corresponds to the infinite temperature limit.
Thus, we can write $G(\mathbf{p})=E(\mathbf{p})+ T k_B D_{\mathrm{KL}}(\mathbf{p},\mathbf{p}^\circ)$,
and interpret the Boltzmann distibution $p_a(T)$ as the distribution that minimizes the energy subject to the normalization constraint
and the constraint on the KL divergence from the uniform distribution.

Now consider the ridge regression problem.
It consists in determining the values of the parameters $a_k$ that minimize the error of the fit $E(\mathbf{a})$, which is analogous to an energy,
under a constraint $H(\mathbf{a})$ on the scale of the parameters (Eq.(\ref{eq:scale})), which is analogous to the normalization condition on the probabilities $\sum_k p_k=1$,
and a constraint on the divergence $F(\mathbf{a},\mathbf{a}^\circ)$ from reference parameters $\mathbf{a}^\circ$ (Eq.(\ref{eq:F})),
which is analogous to the divergence $D_{\mathrm{KL}}(\mathbf{p},\mathbf{p}^\circ)$ from the uniform distribution.
This formal analogy suggests that the ridge parameter $\Lambda$ (or $(1-\mu)\Lambda$, with our definition of the rescaled ridge regression problem in Eq.(\ref{eq:scaled}))
plays a role analogous to that of the temperature in a statistical mechanical system.
It is therefore interesting to compare the behaviour of the two systems in the zero temperature and in the infinite temperature limit.

In the zero temperature limit, only the microstate of minimum energy contributes to the Boltzmann distribution
of a thermodynamic system. We can interpret this state as the state dominated by the correlations between
degrees of freedom embodied in the energy function.
In the ridge regression context, the contribution to the fit of each principal component (i.e. each eigenvector $\mathbf{u}^\alpha$ of the correlation matrix)
is weighted by the factor $1/(\lambda_\alpha + \Lambda)$ (Eqs.(\ref{eq:A_spectral},\ref{eq:Asca})),
so that the relative contribution of the eigenvector $\mathbf{u}^1$ corresponding to the minimum eigenvalue $\lambda_1$
is maximal when $\Lambda=0$.
The similarity with a thermodynamic system is particularly striking if $\lambda_1 \ll \lambda_2$ as, in that case,
the eigenvector $\mathbf{u}^1$ is the only one that significantly contributes to the regression at $\Lambda=0$.

On the other hand, in the infinite temperature limit, the probabilities $p_k$ tend to the reference values $p^\circ_k = 1/K$,
i.e. the distribution no longer depends on the energies and all microstates are equally populated.
Similarly, when $\Lambda\rightarrow\infty$ in ridge regression, the parameters $a_k$ tend to their reference values $a^\circ_k$
($a^\circ_k=0$ in non-scaled regression, and $a^\circ_k = a\sca_{k,\infty}$ if we set $\xi=\xi_\infty$ or $\xi=\xi_\Lambda$ in rescaled regression, Eq.(\ref{eq:ascinf})),
which no longer depend on the correlation matrix, except for the scaling factor.
Furthermore, the equipopulation of the thermodynamic microstates finds an interesting correspondance in the fact that
the factors $1/(\lambda_\alpha + \Lambda)$ weighting the contributions of the eigenvectors $\mathbf{u}^\alpha$ to the fit are all equal when $\Lambda \gg \lambda_\alpha$.

To further analyze this analogy, we can compute the derivative of the error of the fit with respect to $\Lambda$, which is equivalent to the specific heat at constant volume $c_V$,
\begin{equation}
c_V=\frac{\partial E}{\partial \Lambda}
=2\Lambda
\frac{\sum_\alpha\frac{(y^\alpha)^2}{(\lambda_\alpha+\Lambda)}}
{\left(\sum_\alpha\frac{(y^\alpha)^2\lambda_\alpha}{(\lambda_\alpha+\Lambda)^2}\right)^2}
\left[
\sum_\alpha\frac{(y^\alpha)^2}{(\lambda_\alpha+\Lambda)}
\sum_\alpha\frac{(y^\alpha)^2}{(\lambda_\alpha+\Lambda)^3}-
\left(\sum_\alpha\frac{(y^\alpha)^2}{(\lambda_\alpha+\Lambda)^2}\right)^2
\right]>0\, .
\label{eq:cV}
\end{equation}
The Cauchy-Schwartz inequality allows to prove that $c_V$ is positive for $\Lambda>0$,
which means that the error of the fit increases with $\Lambda$ for $\Lambda>0$,
as is intuitive since $\Lambda$ imposes a constraint that is fulfilled at the cost of increasing the error.
Thus, the positivity of $c_V$ for $\Lambda > 0$ provides additional support to the thermodynamic analogy.

\subsection*{Maximum penalty fit}

We now propose two criteria for choosing suitable Tykhonov parameters inspired by the above statistical mechanical analogy.
A good value of $\Lambda$ should provide a satisfactory trade-off between minimizing the error and reducing overfitting.
In the small $\Lambda$ phase the error is small but $F$ is large and overfitting is the main problem, while in the large $\Lambda$ phase the contrary holds.
We conjecture that, if these phases are separated by a second order phase transition with diverging specific heat in the limit of infinite variables, we may observe a peak of the specific heat for the actual number of variables $P=11$.
Thus, we define $\Lambda_{c_V}$ as the value of $\Lambda$ at which the specific heat has a maximum.
There is only one such value $\Lambda_{c_V}>0$, which can be numerically determined by maximizing the specific heat Eq.(\ref{eq:cV}).

The second criterion that we propose is based on the ``entropic'' contribution to the free energy, $(1-\mu)\Lambda F$.
This quantity is equal to zero both at $\Lambda=0$, where all the information of the correlation matrix is retained,
and for $\Lambda\rightarrow\infty$, where $\mathbf{a}=\mathbf{a}^\circ$ and thus $F=0$, and the information of the correlation matrix is lost (except for the scaling factor).
In between, the penalty $(1-\mu)\Lambda F$ reaches a maximum, and we hypothesize that this maximum corresponds to a possibly optimal choice of the ridge parameter $\Lambda$.
We call ``maximum penalty (MP) fit'' ridge regression with this choice of $\Lambda$.

We consider the case of the rescaled ridge regression, with the reference parameters chosen equal to the parameters obtained in the $\Lambda\rightarrow\infty$ limit,
i.e. $\xi=\xi_\infty$ and $\mathbf{a}^\circ=\mathbf{a}\sca_\infty$.
Using Eqs.(\ref{eq:Asca}-\ref{eq:F}), the MP ridge parameter $\Lambda_\mathrm{MP}$ is then defined as the value of $\Lambda$ that maximizes the term
\begin{equation}
(1-\mu)\Lambda F=(1-\mu)\Lambda\sum_\alpha y_\alpha^2\left(\frac{\nu(\Lambda)}{\Lambda+\lambda_\alpha}-\xi_\infty\right)^2\, .\label{lambda_MP}
\end{equation}
Since $\nu(\Lambda)=\Lambda\xi + (1-\mu/2)/(1-\mu)$, the $(1-\mu)$ factor can be expressed as follows:
 $1-\mu=1/\left(1+2\Lambda \left(\xi(\Lambda)-\xi_\infty\right)\right)$,
with $\xi(\Lambda)=(\nu-1)/\Lambda=\sum_\alpha \frac{(y_\alpha)^2}{(\Lambda+\lambda_\alpha)^2}/\sum_\alpha \frac{\lambda_\alpha(y_\alpha)^2}{(\Lambda+\lambda_\alpha)^2}$.
Note that the dependency of $(1-\mu)$ on $\Lambda$ actually has only a very minor effect on the determination of $\Lambda_\mathrm{MP}$.
Therefore, a possible approximation is to simply maximize $\Lambda F$, considering $\mu$ as a constant.
We have verified that it yields highly similar, although slightly reduced, values of $\Lambda_\mathrm{MP}$.
We tested two additional alternative definitions of $\Lambda_\mathrm{MP}$, based on different choices of the reference parameters $\mathbf{a}^\circ$,
by setting either $\xi=\xi_\Lambda$ or $\xi=0$ instead of $\xi=\xi_\infty$.
We found that these choices produce results that are generally similar,
although a bit poorer on average, than those based on the above definition.
More detail, and comparative plots presenting the performances of these alternative definitions are given in Supp. Text S1 and Supp. Fig. S1.

The new criteria presented above are compared with two well-known schemes for selecting the Tykhonov parameter,
namely the Range Risk (RR) optimization \cite{RR} and the Generalized Cross Validation \cite{GCV}.
The GCV criterion is based on the minimization of the error of the fit with a penalty on the effective number of fitted parameters, which results in minimizing the following quantity:
\begin{equation}
GCV(\Lambda)=\frac{\frac{1}{N}E(\Lambda)}
{\left[1-\frac{P}{N}+\frac{1}{N}\Lambda\sum_\alpha \frac{1}{\lambda_\alpha+\Lambda}
\right]^2}
\label{eq:GCV}
\end{equation}
where $E(\Lambda)=\sum_\alpha (y^\alpha)^2
\left(1-\frac{\lambda_\alpha+2\Lambda}{(\lambda_\alpha+\Lambda)^2}\right)$ is
the error of the fit, $N$ is the number of samples and $P$ is the number of parameters.
Although GCV may be defined for negative $\Lambda$ as well, we only considered positive $\Lambda$ since the objective function diverges at $\Lambda=-\lambda_\alpha$,
which causes numerical instabilities, and since we observed that the performances of the GCV fit generally worsen if negative $\Lambda$ are allowed.
The RR criterion leads to a related formula.
Nevertheless, we found that these two schemes produce the same results up to numerical precision in the examined case, and we discuss only GCV.  

Finally, we consider the traditional two-parameters fit of B-factors, i.e.
$\mean{\left|\Delta \vec{r}_i\right|^2} \approx$ $a\pr_{\mathrm{ENM}} \mean{\left|\Delta \vec{r}_i\right|^2}_{\mathrm{ENM}} + a\pr_{\mathrm{trasl}} $.
We call this fit the ``no-rotation'' (NoRot) fit, since it neglects all variables associated with rigid rotations,
while the intercept of the fit can be interpreted as the contribution of rigid-body translations.
We also consider the one-parameter fit $\mean{\left|\Delta \vec{r}_i\right|^2} \approx$ $a\pr_{\mathrm{ENM}} \mean{\left|\Delta \vec{r}_i\right|^2}_{\mathrm{ENM}}$,
which neglects all rigid-body degrees of freedom (hence we refer to it as the NoRigid fit).
This fit has the advantage that the fitted force constant $1/a\pr_{\mathrm{ENM}}$
is guaranteed to be positive if the covariance between the experimental B-factors and those predicted through the ENM is positive, which is always the case in our data sets.

\section*{Numerical results}

\begin{figure}[!ht]
\centerline{
\includegraphics[width=0.75\linewidth]{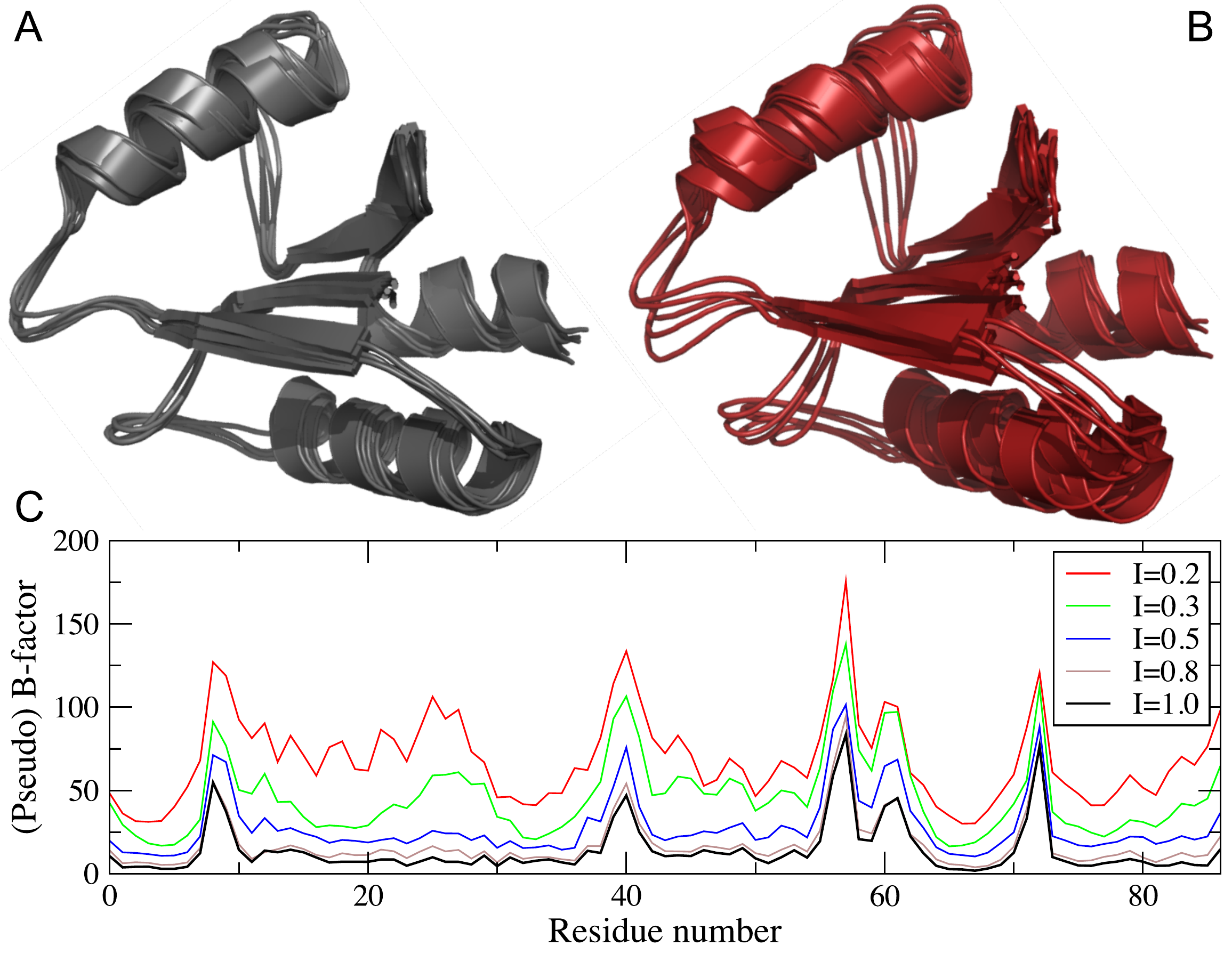}
}
\caption{{\bf Illustration of the thermal fluctuations due to internal and rigid-body degrees of freedom.} (A) NMR ensemble of bacteriophage T4 glutaredoxin, code PDB 1de1. (B) The corresponding simulated ensemble, in which the fraction of motions due to internal degrees of freedom $I=0.2$, and $T=R=0.4$. For clarity, only five structures in each ensemble are shown. (C) The pseudo B-factors for the same protein, as of function of the position in the sequence, in the NMR ensemble 1de1 ($I=1.0$), and in the simulated sets with different values of $I$, with $T=R$.}
\label{fig:1de1}
\end{figure}

The fitting procedures described above were applied to 35 different protein datasets.
The X-ray dataset consists of 376 monomeric proteins with experimentally determined B-factors, and corresponds to the real case application.
The NMR dataset consists of 183 monomeric proteins for which pseudo B-factors were computed from the structural variability within the NMR ensemble.
In this case, the superposition of the structures in each ensemble ensures that all rigid-body motions are excluded from the thermal fluctuations,
and that the pseudo B-factors are representative of internal motions only.
In addition, we created $N_s=33$ sets of simulated data, by adding randomly generated rotations and translations to the internal fluctuations present in each NMR ensemble,
in such a way that the average fraction of motion due to internal, rotational, and translational degrees of freedom is $I_{s}$, $T_{s}$ and $R_{s}$, respectively, in each set $s$ (see Methods).
An example of the atomic fluctuations for a protein in the NMR dataset and in simulated sets with rigid-body motions of increasing amplitude is given in Fig.\ref{fig:1de1}.
Interestingly, even though the overall shape of the (pseudo) B-factor profiles remain very similar with low to medium amplitudes of added rigid-body motions,
critical alterations such as the formation of additional peaks can be observed with larger amplitudes of added rotations (e.g. with $I\leq0.3$ in Fig.\ref{fig:1de1}C).

For each protein, the predicted fluctuations due to internal motions were obtained with the TNM program \cite{TNM},
and were used to estimate the force constants $\kappa$ and the fractions of degrees of freedom $I$, $T$, and $R$, for each protein in each dataset (see Methods).
Besides the force constant, two parameters define the force field of the TNM, the exponent $E$ and the cutoff $C$.
We adopt here the parameters $E=6$ and $C=4.5$\AA, as these values produce optimal predictions of interatomic distance fluctuations (unpublished).
We briefly discuss later the impact of choosing different values of these parameters.

\subsection*{Strength of regularization}

\begin{figure}[!ht]
\centerline{
\includegraphics[width=0.9\linewidth]{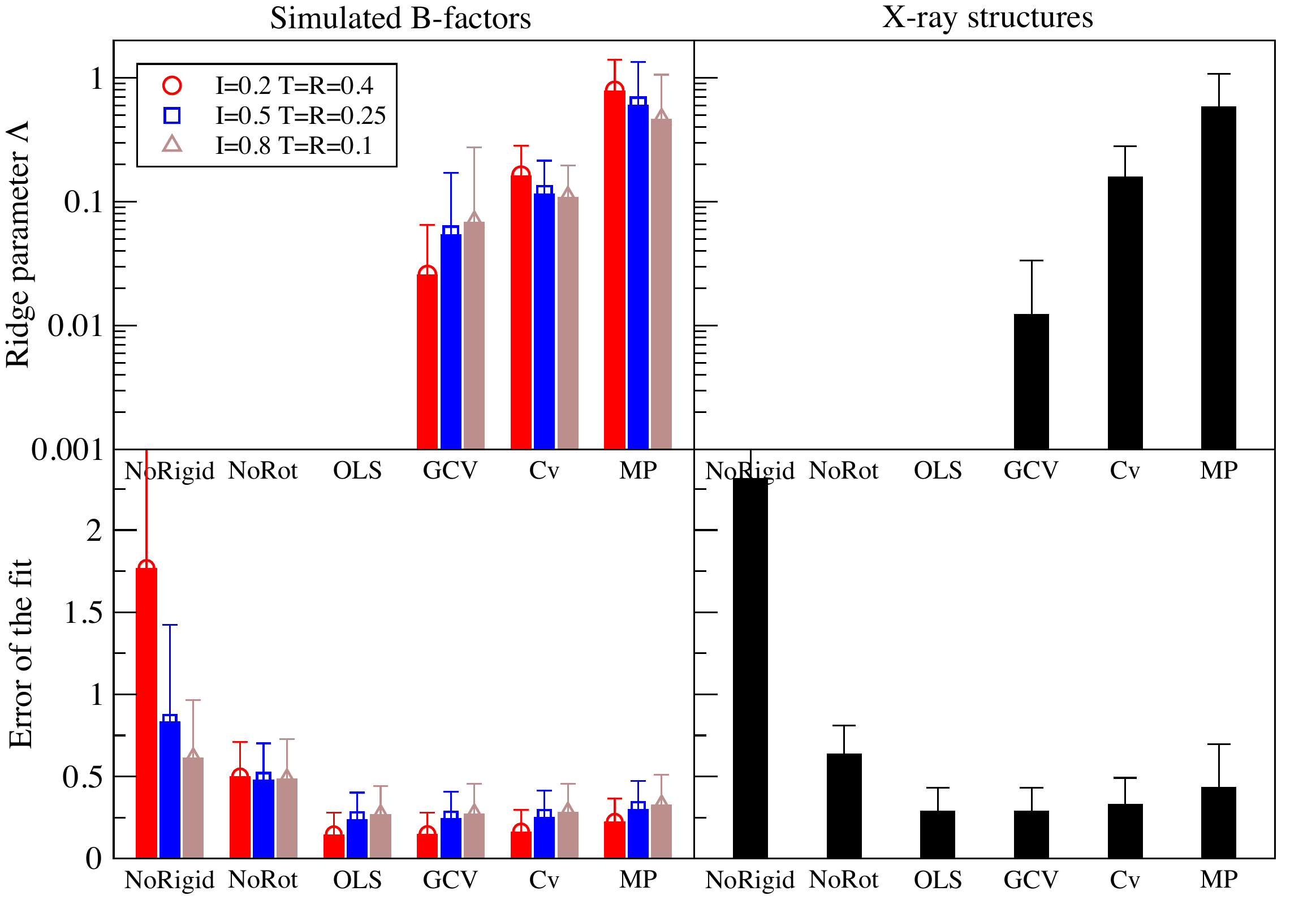}
}
\caption{{\bf Strength of the regularization associated with different types of fit of the B-factors.}
Top: Average value of the ridge parameter $\Lambda$ on selected sets of simulated data (left) and in the X-ray dataset (right).
Bottom: relative error of the fit (Eq. \ref{eq:errorFit}). The error bars correspond to the standard deviation over all proteins in each set.}
\label{fig:lambda}
\end{figure}

The average values of the ridge parameter $\Lambda$ and of the error of the fit are reported on Fig.\ref{fig:lambda},
for different datasets and fitting procedures.
More precisely, we compare the ordinary least square regression (OLS) with
the rescaled ridge regression using different regularization criteria:
Generalized Cross Validation (GCV), specific-heat maximum (Cv), and Maximum Penalty (MP).
Furthermore, the two-parameters fit that neglects rigid-body rotations is referred to as NoRot,
and the one-parameter fit that neglects both rigid-body rotations and translations as NoRigid.

Overall, the comparison of the respective behaviour of these different fits
is very consistent across all datasets, which does support the validity of the NMR data with added rigid-body motions as relevant simulated test sets.
The GCV criterion generates very small values of the ridge parameter $\Lambda$, e.g. $0.01$ on average in the X-ray dataset (Fig.\ref{fig:lambda}, top).
This is due to the fact that the number of data points $N$ (i.e. the number of atoms) is much larger than the number of parameters $P=11$, so that the effective number of parameters (eq. \ref{eq:GCV}) depends very weakly on $\Lambda$ and is always approximately equal to $P$.
The results produced by the GCV fit are therefore very close to those obtained by minimizing the error of the fit without any regularization, as in OLS regression ($\Lambda=0$).
The new criteria that we introduce here, Cv and MP, both yield substantially larger values of $\Lambda$, on all examined datasets.
For example, in the X-ray dataset, the average $\Lambda$ is equal to $0.16$ with the Cv criterion, and to $0.59$ with the MP criterion.  
With larger values of $\Lambda$, the constraint imposed on the parameters of the fit becomes stronger, resulting in somewhat larger fitting errors (Fig.\ref{fig:lambda}, bottom).
However, even with the MP fit that imposes the strongest regularization, the error remains lower than with the common two-parameters fit (NoRot).
On average over all simulated sets, the relative error of the fitted B-factors decreases from $0.50$ with the NoRot fit, to $0.28$ with the MP fit, and $0.21$ with the OLS fit.
The corresponding values in the X-ray dataset are $0.64$ with the NoRot fit, $0.44$ with the MP fit, and $0.29$ with the OLS fit.

\subsection*{Optimal criterion for selecting the ridge parameter}

In the simulated datasets, the contribution of internal degrees of freedom to the fluctuations of the atomic coordinates is known exactly.
These sets give thus the possibility to assess and compare the quality of the various fitting schemes and regularization criteria.

\begin{figure}[!ht]
\centerline{
\includegraphics[width=0.55\linewidth]{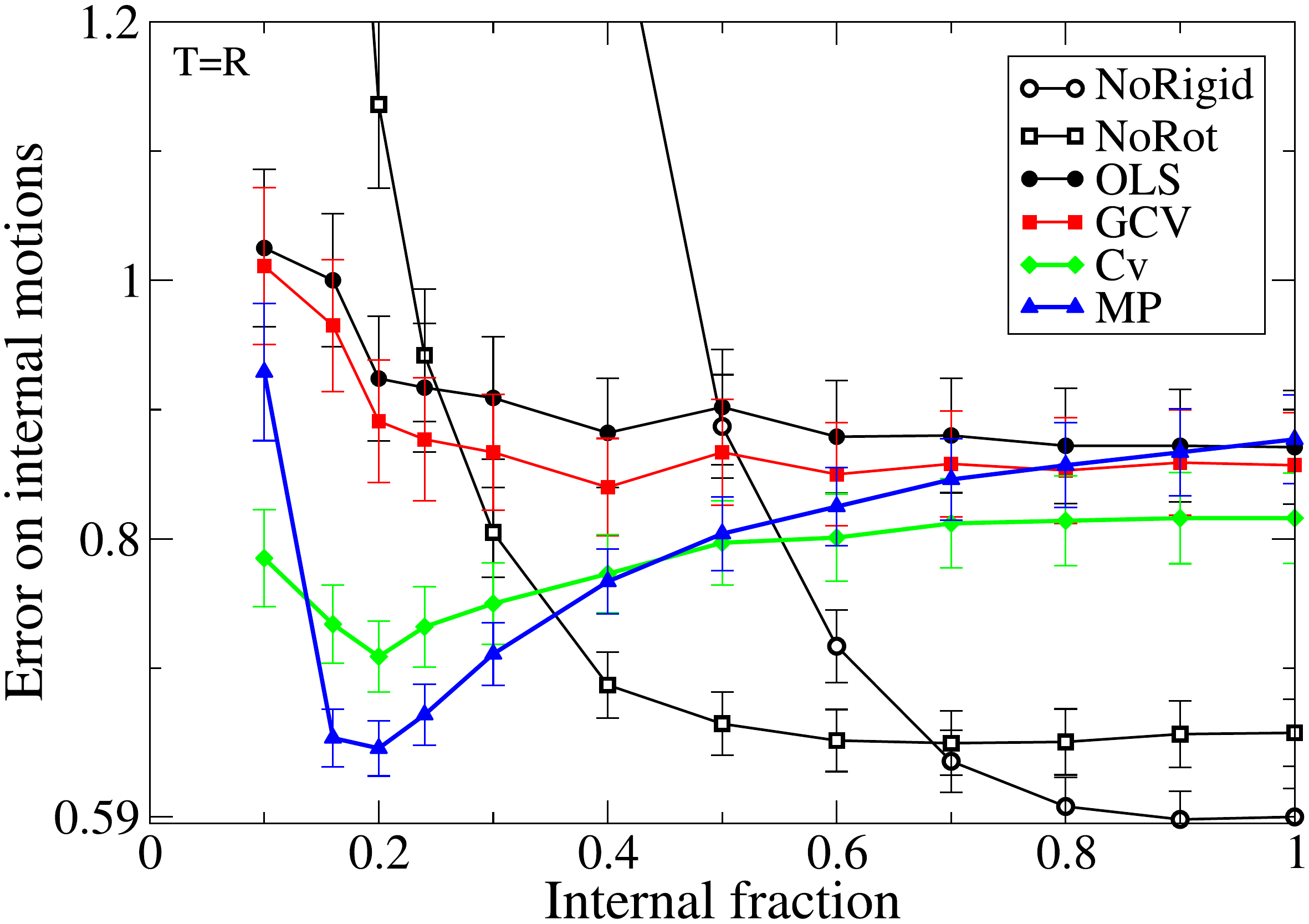}
}
\caption{{\bf Error of the fitted internal motions.} $E^{\mathrm{int}}$ (Eq. \ref{eq:errorInt}) is given as a function of the fraction of internal motions $I$, in the simulated sets with $T=R$, for different types of fit.
The lowest possible value of the error, $E^{\mathrm{int}}=0.59$, is obtained with the NoRigid fit on the NMR dataset ($I=1.0$).}
\label{fig:errint}
\end{figure}

The error on internal motions $E^{\mathrm{int}}$ (Eq. \ref{eq:errorInt}) reflects the ability to accurately retrieve the fluctuations due to internal degrees of freedom
from a fit of B-factor data that may also include contributions from rigid-body degrees of freedom.
This error arises in part from the imperfections of the elastic network model used to make the predictions,
and in part from the presence of ``noise'' in the B-factor data, in the form of fluctuations due to rigid-body motions.
Since we are here interested in the latter source of error,
we take as reference the lowest possible value of this error ($E^{\mathrm{int}}=0.59$) that is obtained with the NoRigid fit, which does not account for rigid-body motions,
on the NMR dataset with $I=1$, which does not contain fluctuations due to rigid-body motions.
The performances of the various fits as a function of the internal fraction $I$ (with $T=R$) are given in Fig.\ref{fig:errint}.
With the NoRigid fit, all fluctuations are interpreted as being due to internal motions, and the error $E^{\mathrm{int}}$ thus rapidly increases as the internal fraction drops.
The NoRot fit, which accounts for internal motions and translations, but not for rotations, presents a quite similar behaviour
except that it is more robust to the addition of rigid-body fluctuations.
It appears therefore as optimal in the presence of rigid-body motions of medium amplitude (i.e. in the range $0.35\leq I\leq 0.7$).

If there are rigid-body fluctuations of larger amplitude ($I\leq0.35$), the NoRot fit is no longer sufficient, and accounting for all degrees of freedom in the fit becomes a critical necessity.
For that purpose, the Cv fit and the MP fit, based on the new criteria that we introduced, both appear clearly superior to the OLS and the GCV fit.
The Cv fit outperforms the OLS and GCV fits in all of the datasets that we examined, with an error on internal motions that 
remains relatively stable across the whole range of $I$ ($0.71\leq E^{\mathrm{int}} \leq 0.82$).
The error of the MP fit is more dependent on the internal fraction $I$,
but it still outperforms the OLS and GCV in most datasets, except for very low amplitudes of added rigid-body fluctuations ($I\geq0.77$).
In the range $0.15\leq I\leq 0.35$, the MP criterion stands out as the optimal choice, yielding a significantly lower error than all other fits. 
In particular, at $I=0.2$, the error on internal motions $E^{\mathrm{int}}$ is as low as $0.64$ with the MP fit,
which is quite impressive considering that the minimum error possibly achievable is $0.59$,
and that the addressed problem is quite challenging since the amplitude of the ``noise'' (rigid-body fluctuations)
is here four times larger than that of the ``signal'' (internal fluctuations).
For comparison, at $I=0.2$, $E^{\mathrm{int}}=0.92$ without regularization (OLS), and $E^{\mathrm{int}}=0.89$  with the GCV criterion.

We also examined the error $E^{\mathrm{int}}$ on the simulated sets with either $T=0$ or $R=0$.
The results are similar to those obtained with $T=R$, except for the NoRot fit (Supp. Fig. S2).
As it could be expected, the NoRot fit performs much better when there are added translations but no rotations,
and much worse when there are added rotations but no translations.
In addition, a related performance measure can be obtained by considering as a proxy of the real force constant
the force constant determined by the NoRigid fit on the NMR set without rigid-body motions, $\kappa^\circ$.
In each simulated set $s$, the comparison between $\kappa^\circ$ and the force constants $\kappa_{s}$ estimated from the various fitting procedures,
leads to the definition of the error on the force constant, $E^{\kappa}$ (Eq. \ref{eq:errorKappa}).
For all fits, the behaviour of $E^{\kappa}$ as a function of $I$ is strongly related to that of $E^{\mathrm{int}}$, and leads to similar conclusions (Supp. Fig. S3).

\begin{figure}[!ht]
\centerline{
\includegraphics[width=\linewidth]{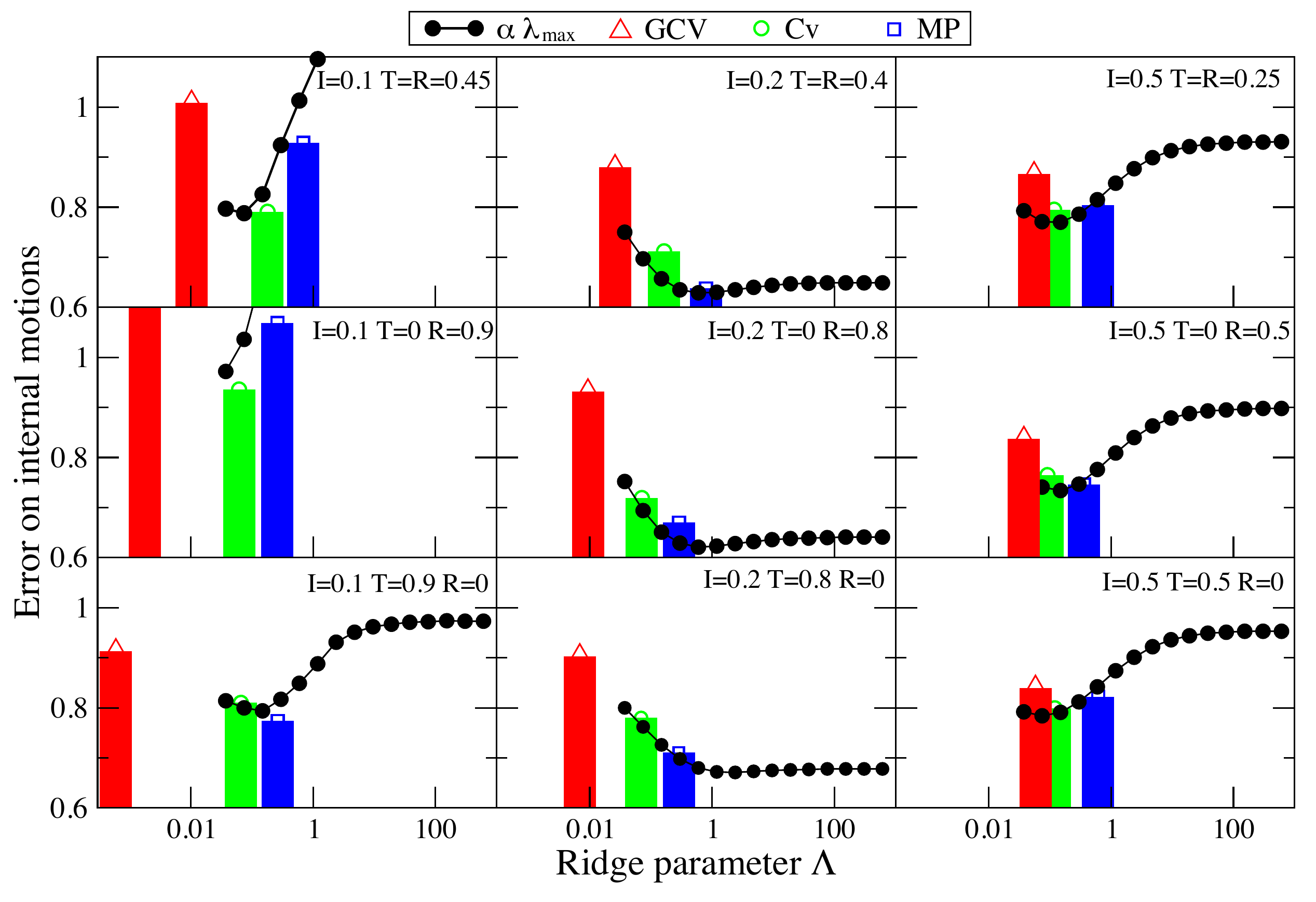}
}
\caption{{\bf Influence of the regularization on the error of the fitted internal motions.} $E^{\mathrm{int}}$ (Eq. \ref{eq:errorInt}) is given as a function of the average ridge parameter $\Lambda$ for different types of fit. Each subplot corresponds to a different simulated dataset, with fractions of degree of freedom: $I=0.1, 0.2, 0.5$ and $T=R$, $T=0$, or $R=0$.}
\label{fig:errint_2}
\end{figure}

Furthermore, to investigate more systematically the impact of using different levels of regularization,
we examined the parametrization $\Lambda=\alpha \lambda_{\mathrm{max}}$,
where $\lambda_{\mathrm{max}}$ is the maximum eigenvalue of the normalized covariance matrix,
and $\alpha$ is a factor that was exponentially increased from $0.02$ to $164$.
We can see in Fig.\ref{fig:errint_2} that for each simulated set, there is an optimal value of $\alpha$,
corresponding to a minimum of the error $E^{\mathrm{int}}$. 
Both the depth and the position of this minimum depend on the fractions of $I$, $T$, and $R$ that define each simulated set.
Yet, in all cases, either the MP or Cv criterion, or both, yield an average $\Lambda$ value and error $E^{\mathrm{int}}$
that are very similar to those obtained at the minimum. 
These results suggest that, for the current application, the new criteria that we introduced not only outperform the GCV criterion,
but are also generally close to optimal in terms of selecting the right level of regularization.

\subsection*{Relative importance of internal and rigid-body motions in crystallographic B-factors}

As detailed above, the analysis of the results obtained on the simulated datasets have shown that the performances of the different
fitting procedures depend to some extent on the fractions of motion due to internal degrees of freedom ($I$),
rigid-body translations ($T$), and rigid-body rotations ($R$).
Notably, if the contribution of rigid-body motions remains small or medium,
accounting for all degrees of freedom may not be necessary, and the use of the NoRigid or NoRot fit can be appropriate.
On the contrary, if rigid-body motions account for a sufficiently high fraction of the fluctuations,
then a full fit including contributions from both translations and rotations is necessary,
and the regularization with the MP criterion yields the best performances. 
Hence, an important question that arises concerns the actual values of the $I$, $T$, and $R$ fractions in the B-factor data from X-ray experiments.
To answer that question, we used the simulated data to evaluate the ability of the different fits to accurately estimate the $I$, $T$, $R$ fractions.

\begin{figure}[!ht]
\centerline{
\includegraphics[width=0.95\linewidth]{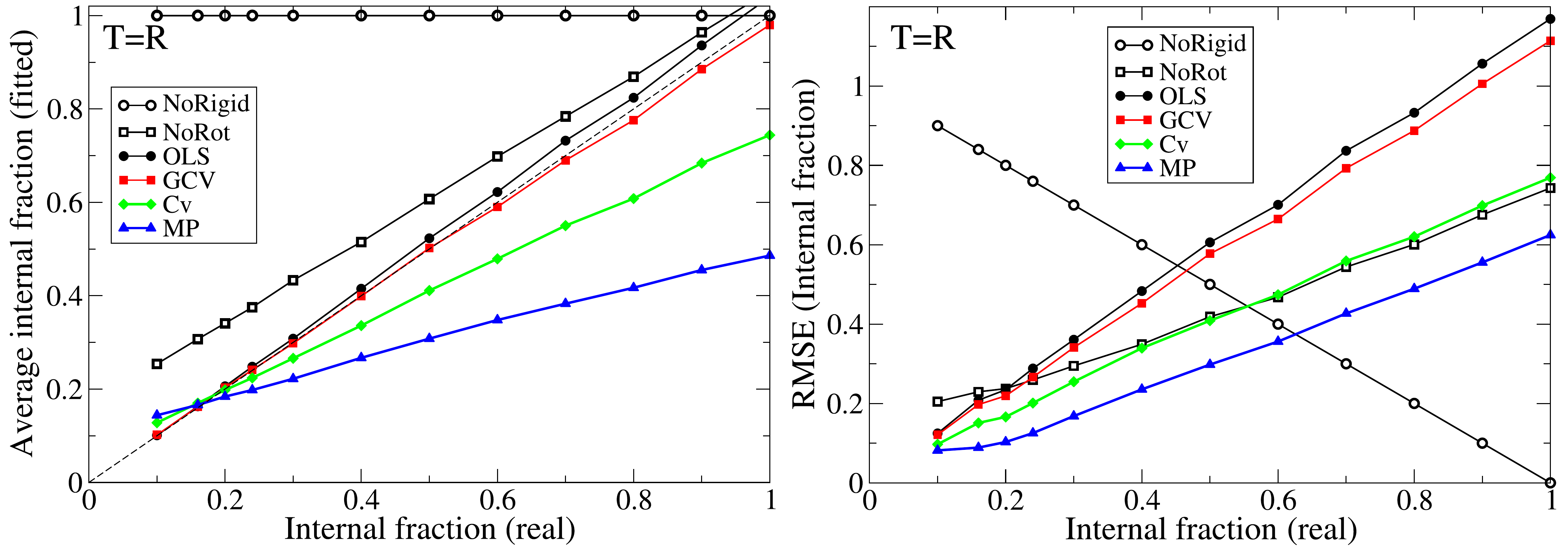}
}
\caption{{\bf Estimation of the fraction of fluctuations due to internal motions, in the simulated datasets.} (left) Average fitted internal fraction $\mean{I_p}$ as a function of the real internal fraction $I$. The dashed line corresponds to $\mean{I_p}=I$. (right) Root mean square error of the fitted internal fraction, RMSE($I_p$) (Eq. \ref{eq:RMSEI}), as a function of $I$.}
\label{fig:ITR}
\end{figure}

The left panel of Fig.\ref{fig:ITR} shows the average value (over all proteins $p$ in a simulated set) of the fitted internal fraction, $\mean{I_p}$, as a function of the actual value of $I$ in the set.
A very strong linear correlation is observed between $\mean{I_p}$ and $I$, for all fits except the NoRigid fit,
which only accounts for internal motions and thus always yields $\mean{I_p}=1$.
The NoRot fit tends to systematically overestimate the importance of internal motions, i.e. $\mean{I_p}>I$, since fluctuations due to rigid-body rotations are
not accounted for and mostly interpreted as being due to internal degrees of freedom.
On the other hand, in fits with $\Lambda>0$, the contribution of internal motions is underestimated when $I$ is large,
and overestimated when $I$ is small (with a threshold at $I\approx0.17$ for all types of fit).
This bias increases with $\Lambda$, and is thus most visible in the MP fit, and practically non-existent in the OLS and GCV fits.

The right panel of Fig.\ref{fig:ITR} shows the root mean square error of the fitted internal fraction, RMSE$(I_p)$, as a function of $I$.
Interestingly, despite the fact that the MP fit is subject to the strongest regularization, and thus affected by the strongest bias $(\mean{I_p}-I)$,
it does yield the lowest error on the estimation of the internal fraction for individual proteins (at least when $I\leq0.6$, otherwise the NoRigid fit is superior).
This is explained by the fact that the bias is accompanied by a sizeable reduction of the variance (between proteins) of the fitted fraction $I_p$.
On the contrary, the OLS and GCV fits are affected by a large variance of the fitted internal fraction.
Therefore, even though there is little to no systematic bias, the error on $I_p$ is actually much larger with these two fits. 
The NoRot and Cv fits present intermediate performances. Similar results hold for the other types of degrees of freedom, $T$ and $R$ (Supp. Fig. S4).

\begin{figure}[!ht]
\centerline{\includegraphics[width=0.9\linewidth]{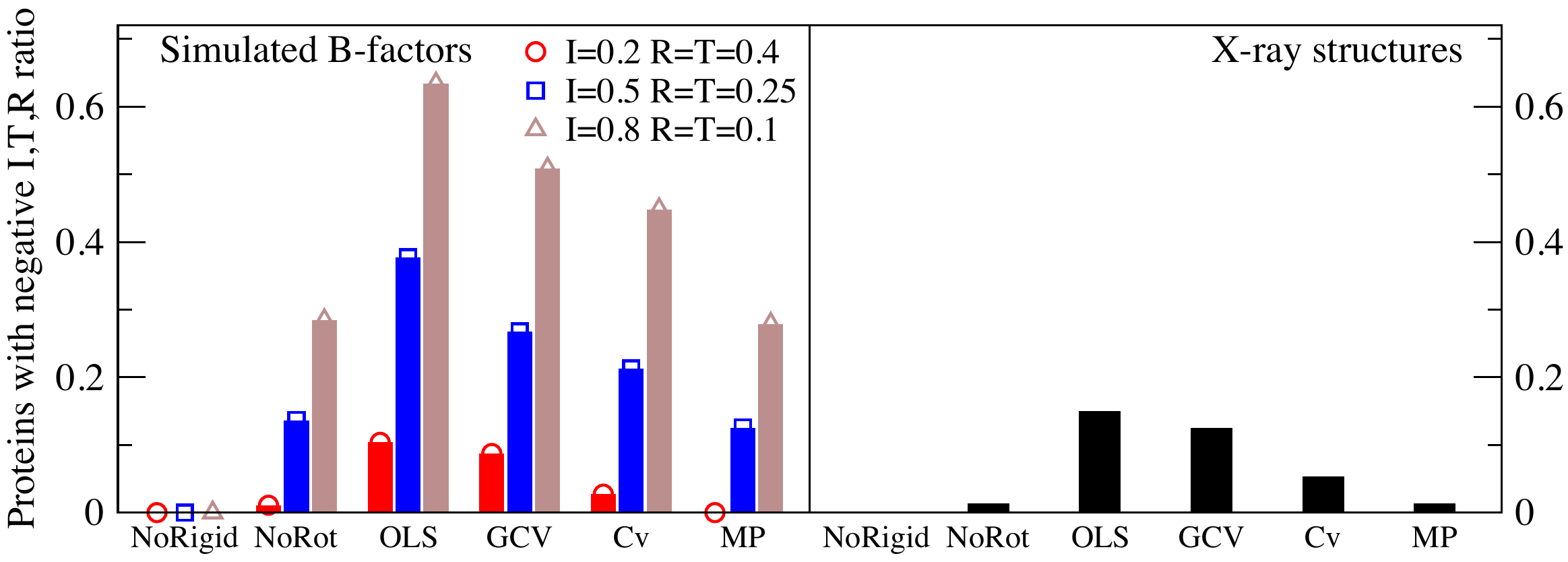}
}
\caption{{\bf Occurrence of unphysical fit parameters.} The fraction of proteins for which the fit produces a negative $I_p$, $T_p$, or $R_p$ ratio is given for the different fits, on selected simulated datasets (left), and on the X-ray dataset (right).}
\label{fig:negITR}
\end{figure}

In a number of cases, the parameters of the fit correspond to negative values of the fraction of motion
due to either internal, translational or rotational degrees of freedom, which is unphysical (in particular, $I<0$ means a negative force constant).
As shown on Fig.\ref{fig:negITR}, this problem is particularly serious when the internal fraction is large.
For example, at $I=0.8$, unphysical parameters are obtained for as many as 63\% of the proteins, with the OLS fit.
Here again, the benefits of regularization are very apparent: e.g. at $I=0.2$, the number of proteins with negative $I_p$, $T_p$, or $R_p$ fractions
is reduced from 10\% with the OLS fit, to 3\% with the Cv fit, and 0\% with the MP fit.
Similar results are obtained on the dataset of crystallographic B-factors.
In this case, unphysical parameters are derived from the OLS fit for about 15\% of the proteins, but this number drops to 1\% with the MP fit.

In summary, the analysis of the results obtained on the simulated data indicates that the MP fit
produces the best estimation of the $I$, $T$, $R$ fractions, for individual proteins
(unless $I>0.6$, in which case the NoRigid fit may be preferable).
This is achieved at the price of a systematic bias, which is a consequence of the regularization of the parameters.
On the contrary, the OLS and GCV fits are characterized by a large RMSE on the fitted $I$, $T$, $R$ fractions,
and moreover often generate unphysical values of the parameters.
However, these fits are not biased, i.e. on average over a sufficiently large set of proteins,
the estimated fractions of motion due to the different types of degrees of freedom, $\mean{I_p}$, $\mean{T_p}$, and $\mean{R_p}$, are remarkably accurate (Fig. \ref{fig:ITR}, left). 
Thus, although the OLS and GCV fits may not be well suited to any real-case application, this absence of bias is an interesting
feature that can be exploited to evaluate the average contributions of internal and rigid-body motions in crystallographic B-factors.

\begin{figure}[!ht]
\centerline{
\includegraphics[width=0.55\linewidth]{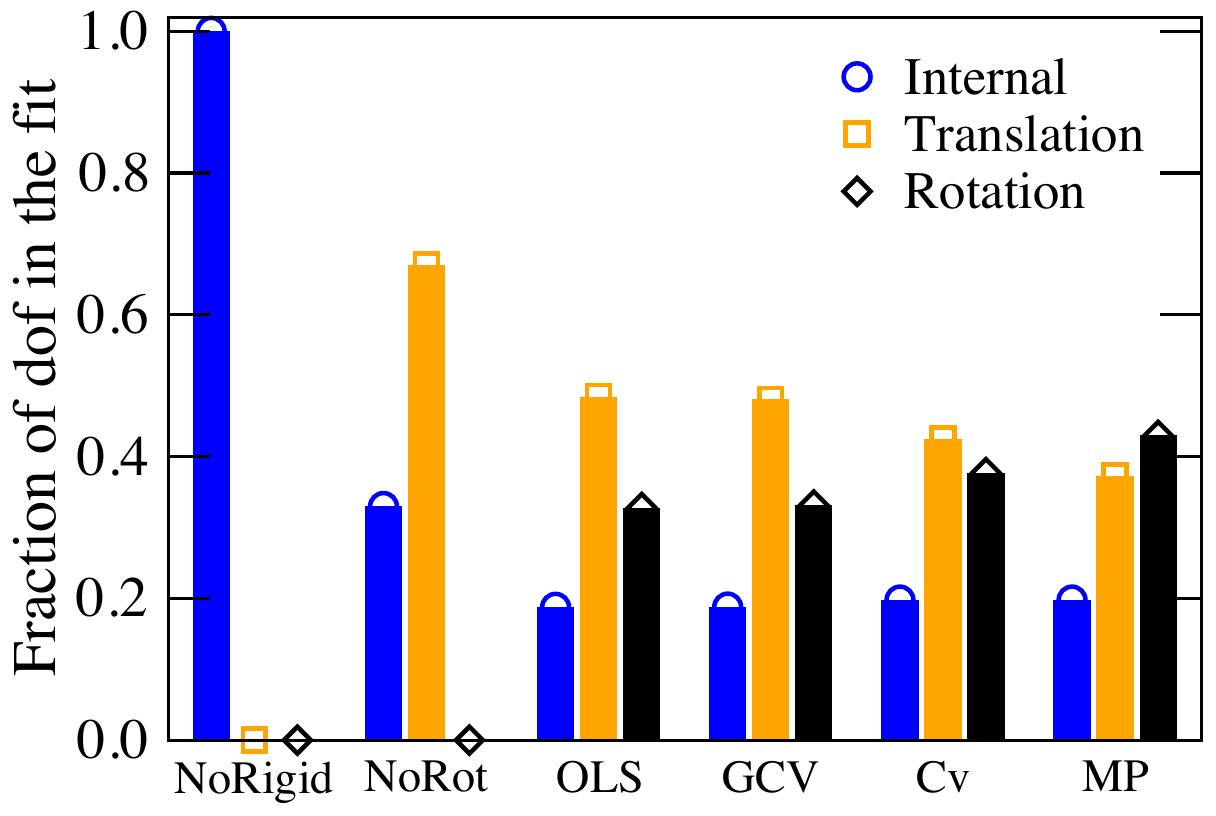}
}
\caption{{\bf Relative contributions of internal and rigid-body degrees of freedom in crystallographic B-factors.} The average fractions of motion estimated from the experimental B-factors are given for the different types of fit. The OLS fit is expected to yield the least biased estimate.}
\label{fig:dof}
\end{figure}

The results for experimental B-factors measured in protein crystals are presented in Fig.\ref{fig:dof}.
The OLS fit, which is expected to give the least biased estimates, yields average values of
$I=0.19$ for internal, $T=0.48$ for translational, and $R=0.33$ for rotational degrees of freedom.
The regularized fits based on the Cv and MP criteria generate very similar values of the fraction of internal motions,
which is consistent with the fact that the bias on $I$ is minimal when $I\approx0.17$ (Fig.\ref{fig:ITR}, left).
These fits do however somewhat underestimate, on average, the contribution of rigid-body translations
and overestimate the contribution of rigid-body rotations (e.g. $T=0.37$ and $R=0.43$ with the MP fit). 
In any case, these results strongly suggest that the contribution of rigid-body motions is quite important in crystallographic B-factors,
with internal motions accounting for 20\% or less of the measured atomic fluctuations.
In this range, the commonly used two-parameters NoRot fit fails to provide satisfactory results,
and the MP fit appears as a much preferable alternative (see e.g. Fig.\ref{fig:errint}).

\subsection*{Variation of force constants}
\begin{figure}[!ht]
\centerline{
\includegraphics[width=0.9\linewidth]{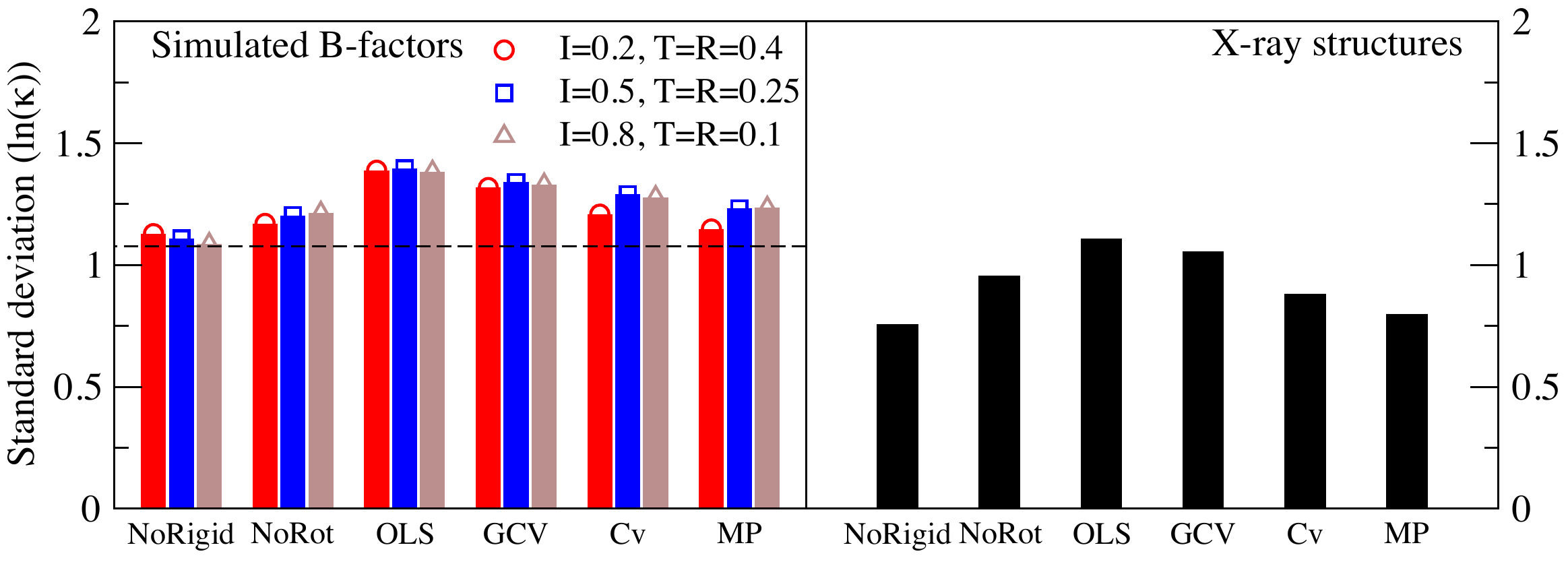}
}
\caption{{\bf Variation of the fitted force constant for different types of fit.} The standard deviation (among different proteins) of the logarithm of the force constant, $\sigma_{\ln(\kappa)}$, is given for selected sets of simulated data (left) and for the X-ray dataset (right).
In each set, the proteins for which at least one of the fits yielded a negative value of $\kappa$ were omitted, for all fits.
The dashed line corresponds to the value of $\sigma_{\ln(\kappa)}$ obtained with the NoRigid fit on the NMR dataset ($I=1$).
A value of $\sigma_{\ln(\kappa)}=1$ corresponds to a multiplicative spread of $e^1\approx2.7$, i.e. for most proteins, the estimated force constant is smaller than $2.7$ times and larger that $(1/2.7)$ times the geometric mean.}
\label{fig:kappa}
\end{figure}

One of the main motivations of this work was to assess how the evaluation of the force constant from a fit of crystallographic B-factors can be improved by properly accounting for fluctuations due to rigid-body degrees of freedom.
As detailed above, this assessment can be performed rigorously with the simulated datasets, in which the relative contributions of internal and rigid-body motions are known \textit{a priori}.
Such knowledge is not available in the X-ray dataset, but it is however possible to evaluate the variability of the force constants estimated by the different fitting procedures across all proteins in the set.
We assess this variability by measuring the standard deviation of the logarithm of the force constant, $\sigma_{\ln(\kappa)}$, plotted in Fig.\ref{fig:kappa}.
We consider the logarithm because its fluctuations are better behaved, and it allows to eliminate the influence of multiplicative scale factors.

In the simulated sets, an important part of the variability is due to the nature of the NMR data.
Typically, the extent of structural variability within an NMR ensemble is determined by the number and quality of the interatomic distance constraints extracted from the experiment, which may depend on a number of factors not directly related to the dynamical properties of the macromolecule.
In consequence, the NoRigid fit applied to the NMR dataset without added rigid-body motions ($I=1.0$) generates
values of the force constants that are already considerably spread out, with $\sigma_{\ln(\kappa)}=1.08$.
The addition of rigid-body motions tends to increase this variability,
although $\sigma_{\ln(\kappa)}$ shows a relatively limited dependence on the internal fraction $I$ (Fig.\ref{fig:kappa}, left).
This is not overly surprising since, in each simulated set, the amplitude of added rigid-body motions is identical for all proteins.
The largest variability amongst force constants evaluated for different proteins occurs with the OLS and GCV fits.
These fits account for rigid-body motions but are not (or only slightly) regularized,
and the collinearity between predictor variables leads thus, in a number of cases, to the determination of either negative or very large values of the force constant.
Increasing $\Lambda$ tends to mitigate this problem, and consequently reduces the variability (Supp. Fig. S5).
The MP fit is thus characterized by an intermediate level of variability of the force constants, similar to that of the NoRot fit.
It is however important to recognize that a low variability does not imply a correct estimation of $\kappa$.
For example, the NoRigid and NoRot fits yield a relatively low variability but systematically underestimate the force constant, since fluctuations due to rigid-body rotations are interpreted as resulting from internal degrees of freedom.

The results are essentially similar in the X-ray dataset (Fig.\ref{fig:kappa}, right).
Here again, we expect some intrinsic variability of the estimated force constants, due to a variety of phenomena that may influence the B-factors measured in protein crystals, such as crystal packing and static disorder \cite{Hinsen2008,Riccardi2009}.
It is therefore unlikely that any fitting procedure could achieve a complete elimination of the spread of the force constants estimated for different proteins.
With $\sigma_{\ln(\kappa)}=0.80$, the MP fit does however produce a substantial reduction of the variability, 
in comparison with both the OLS fit ($\sigma_{\ln(\kappa)}=1.11$), in which case we can interpret the excess variability as due to overfitting,
and the commonly used NoRot fit ($\sigma_{\ln(\kappa)}=0.96$).

\subsection*{Influence of the parameters of the ENM force field}
\begin{figure}[!ht]
\centerline{
\includegraphics[width=0.9\linewidth]{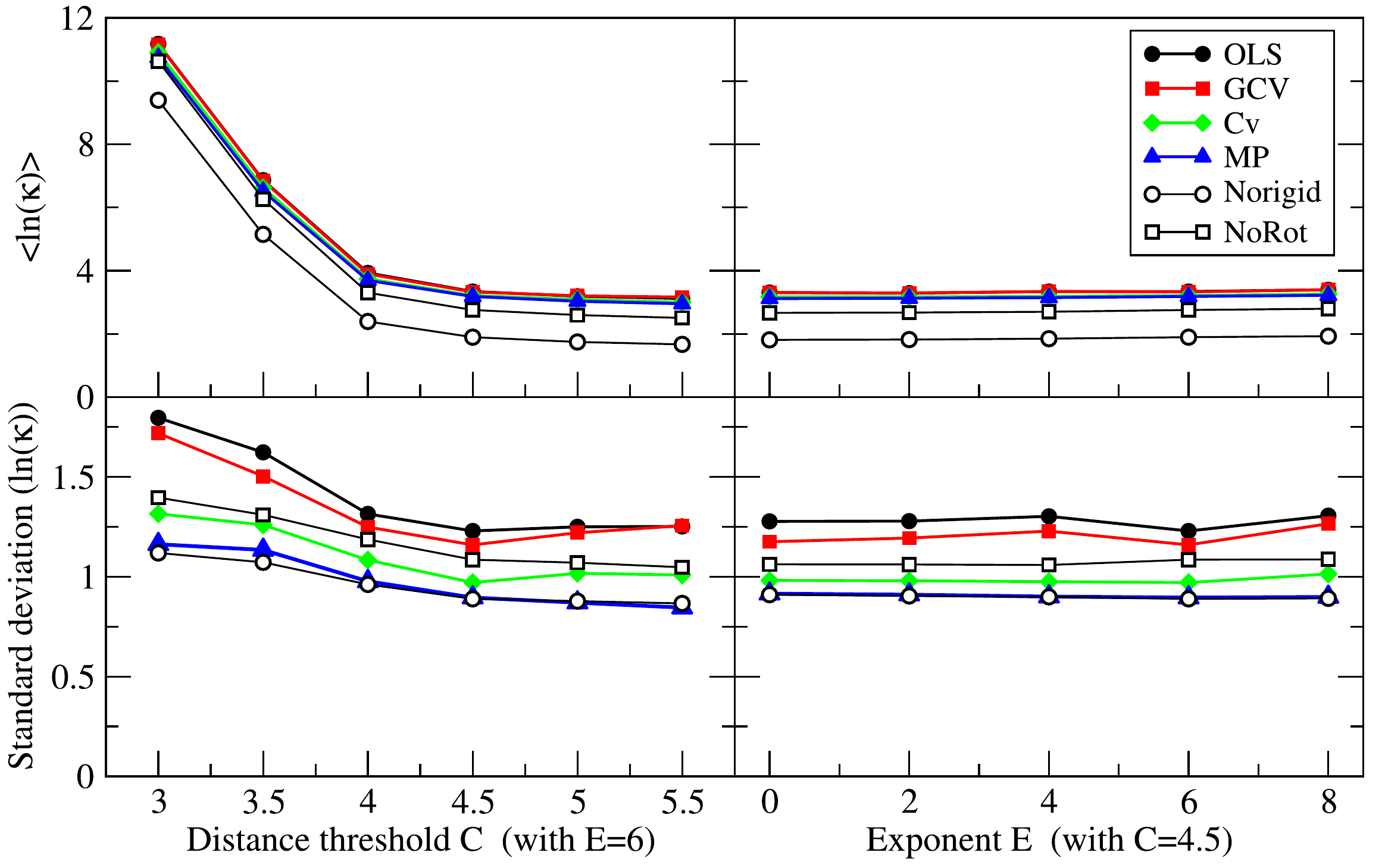}
}
\caption{{\bf Dependence on the ENM parameters.} The average value (top) and the standard deviation (bottom) of the logarithm of the force constant $\kappa$, eliminating proteins for which $\kappa$ is negative, are plotted versus the distance cut-off $C$ for exponent $E=6$ (left) and versus the exponent $E$ for distance cut-off $C=4.5$\AA (right).
}
\label{fig:cut-off}
\end{figure}

In the elastic network model, the stiffness of the spring associated to a given pair of residues is typically defined as a function of the spatial distance
separating these two residues. We adopted here a common expression of the force constant, $\kappa_{ij}=\kappa(r_0/r_{ij})^E$ if $r_{ij}\leq C$, and $\kappa_{ij}=0$ otherwise,
where $r_{ij}$ is the minimal interatomic distance between residues $i$ and $j$, and $r_0$ ($=3.5$\AA) is a reference distance (see Methods).
The results presented above are related to the determination of the factor $\kappa$, with chosen values of the distance threshold $C=4.5$\AA, and of the exponent $E=6$.
In order to analyze the dependence of the force constant on $C$ and $E$ and 
to ensure the robustness of our conclusions, we investigated the influence of these two parameters of the ENM on the estimated force constants.
In practice, we varied $C$ from $3.0$ to $5.5$\AA, and $E$ from 0 to 8, and applied the resulting models to the X-ray dataset.
The average and the standard deviation of the logarithm of the fitted force constants are given in Fig.\ref{fig:cut-off},
for the various fits, with different values of $C$ and $E$.

As could be expected, the average force constant tends to decrease when the distance cut-off $C$ increases (Fig.\ref{fig:cut-off}, top left).
Indeed, if the number of interacting pairs of residues increases due to a larger cut-off, the force constant associated with each pair must decrease to maintain a similar amplitude of the atomic fluctuations.
We also measured the correlation coefficients between the force constants determined at different values of $C$ and $E$,
over all proteins of the X-ray data set, for each type of fit. 
The force constants obtained for the different proteins, with two different sets of parameters $C$ and $E$,
are always highly correlated to each other (correlation coefficient larger than 0.98) as long as $C\geq 4$\AA.
Drastic changes do however occur when the cut-off distance $C$ is decreased below 3.5\AA:
forces constants determined with $C\leq3.5$\AA$ $ tend to present vanishing correlations
with those determined using different values of $C$ and $E$, irrespective of the type of fit.
These observations are consistent with the fact that if the distance cut-off is too small,
crucial interactions are ignored, which can result in a disruption of the overall integrity of the structure, and have dramatic effects on the predicted dynamics.
In contrast, the force constant depends much more weakly on the exponent $E$ (Fig.\ref{fig:cut-off}, right).
This is due to the choice of the reference distance $r_0=3.5$\AA. Indeed, when $E$ increases, $\kappa_{ij}$ increases for pairs of residues with $r_{ij}<3.5$\AA,
and decreases for pairs or residues with $r_{ij}>3.5$\AA. These two effects appear to compensate each other fairly well, which explains the small impact of $E$ on the scaling factor $\kappa$.

Importantly, all types of fits show the same qualitative behavior, and the differences between the fits are mostly independent of the choice of the $E$ and $C$ parameters,
which demonstrates the robustness of the results presented in the previous sections, where we used $C=4.5$\AA$ $ and $E=6$.
Note however that, even if the parameters $E$ and $C$ have a relatively limited effect on the force constant $\kappa$,
they may significantly affect the normal modes of motion predicted by the ENM, and thus the overall performances of the model.

\section*{Conclusions}

Crystallographic B-factors are commonly used to train and validate computational models of protein dynamics.
There are however a number of possible shortcomings to this approach \cite{Riccardi2010,Fuglebakk2012,Fuglebakk2013}.
In particular, rigid-body rotations are often neglected,
even though several studies concluded that the occurrence of such motions is a major determinant of experimental B-factors \cite{Kuriyan1991,Soheilifard2008,Hafner2010,Lezon2012}. 
Our results support these conclusions, and indicate that
the contribution of internal motions is lower than 20\%, on average.
A systematic analysis of simulated sets of pseudo B-factors, characterized by variable amplitudes of rigid-body fluctuations,
suggests that these estimations are subject to very little bias.

These results are also consistent with a study by Halle \cite{Halle2002} and a follow-up by Li and Br\"uschweiler \cite{Li2009},
who showed that B-factors in X-ray protein structures are well predicted by the number of contacts of each residue.
In light of the importance of rigid-body motions to B-factors,
this observation may be explained by the negative correlation between the number of contacts,
which tends to be larger for buried residues, and the atomic fluctuations due to rigid-body rotations, which more severely affect surface residues as they are further away from the center of mass.
On the other hand, the number of contacts also has a strong influence on the fluctuations due to internal degrees of freedom, predicted for example by the ENM.
This entails a high level of collinearity between explanatory variables and underlines the fact that,
to determine the scale of the ENM force constant via a fit of B-factors,
it is not only important to properly account for all rigid-body degrees of freedom,
but also to carefully regularize the fit in order to reduce the risk of overfitting.

For that purpose, we introduced two novel criteria for determining the ridge parameter $\Lambda$.
The definition of these criteria was motivated by a strong formal analogy between ridge regression and statistical mechanics,
where $\Lambda$ plays the role of the temperature.
Our results demonstrate that the MP criterion is an almost optimal way of choosing $\Lambda$, at least for the problem of fitting force constants from B-factors.
For simulated B-factors with an internal fraction close to $I=0.2$ (the range that we expect to find in X-ray structures),
the MP fit was shown to yield the minimum root mean square error (RMSE) on the estimation of the fluctuations due to internal degrees of freedom,
the minimum RMSE on the logarithm of the fitted force constant,
and the minimum RMSE on the estimated fractions of internal, translational and rotational motions.
In contrast, the commonly used GCV criterion produces a 40\% larger value of the RMSE on internal fluctuations.
Furthermore, in the X-ray dataset, the GCV fit generates unphysical values of the parameters (e.g. negative force constants)
for 47 out of 376 proteins, while this number is reduced to 5 with the MP fit.
The GCV criterion also induces an increase of the across-protein variability of the logarithm of the force constants, from $0.80$ with the MP criterion to $1.06$.
The poor performances of the GCV fit indicate that the ridge parameter determined by this criterion is too small and does not
provide a sufficient regularization of the fit of B-factors. This is most likely due to the fact that
overfitting occurs here because of the high level of collinearity between explanatory variables,
even though the number of fitted parameters ($P=11$) remains small with respect to the number of data points ($N=$ number of residues).

The Cv criterion produces a level of regularization that is intermediate between the GCV and MP fits.
Despite yielding poorer performances than MP when the internal fraction is close to $I=0.2$, and thus when applied to X-ray structures,
the Cv criterion is consistently superior to GCV over all sets of simulated data, and it outperforms MP when the internal fraction is either large or small.
In consequence, even though the Cv fit appears to be somewhat less well adapted than MP to the present application,
it has the advantage of being more robust with respect to the nature of the analysed data, and is thus a good candidate for application in a wide range of situations. 

Note that the analogy between statistical mechanics and ridge regression that we explored here can easily be extended to other types of constrained minimization problems,
where the weight of the constraint imposed on the parameters has to be fixed in an optimal way.
In particular, in the context of statistical mechanics our analogy evidences the existence of an intrinsic temperature that maximizes the penalty term $T\left(S(T)-S(T\rightarrow\infty)\right)$,
where $T$ is the temperature and $S$ is the entropy, at which the minimization of the energy and the maximization of the entropy are well balanced.

In summary, the present results further confirm the predominance of rigid-body motions in crystallographic B-factors,
and underline the importance of accounting for all degrees of freedom via a carefully regularized fit
when B-factors are used to scale computational models of protein dynamics.
The MP fit stands out as the optimal choice for that particular application.
Importantly, the potential benefits of the MP and Cv criteria are not limited to the fit of B-factor data.
Although further studies would be necessary to assess the general applicability of these criteria to the regularization of other multi-variable fits,
the strong performances displayed here, in comparison with the common GCV approach,
suggest that the adoption of these new criteria could be advantageous in various applications.
This may be particularly true for problems that bear similarity to the considered case,
i.e. when the number of fitted parameters is small enough with respect to the number of data points but the explanatory variables are highly collinear,
and/or when restrictions apply to the physically acceptable values of the parameters.
On the other hand, the rescaled variant of ridge regression that we introduced here is readily applicable to any regression problem in which the intercept of the fit has a physical meaning and must be penalised similarly to the other explanatory variables.

\section*{Methods}

\subsection*{Data sets}
We examined a test set of 380 non-redundant monomeric proteins whose structure has been solved by X-ray crystallography, with resolution better than 2\AA,
extracted from the Top500 dataset used to benchmark the MolProbity program \cite{MolProbity}.
From this set, we eliminated the proteins with pdb codes 2sns, 1cne, and 2ucz,
because the record of B-factors was missing in these proteins (all recorded B-factors were equal),
and 1rho, which was the most serious outlier for all the fits and was predicted as hexameric by the software PISA \cite{pisa}.

The 183 structural ensembles in the NMR dataset were selected according to the following criteria: they consist of at least 20 models with identical number of residues;
they correspond to monomeric proteins of at least 50 residues that present at most 30\% sequence identity with one another;
they are not listed under the SCOP classifications "Peptides" or "Membrane and cell surface proteins";
they do not include ligands, DNA or RNA molecules, chain breaks, or non-natural amino acids;
and they do not contain highly flexible loops or C- or N-terminal tails.
To enforce the latter criterion, highly flexible regions were defined as stretches of at least two consecutive residues
for which the mean square fluctuations of the C$_{\alpha}$ coordinates are larger than 2.5 times the average over all residues. 
Such loops or tails typically correspond to disordered regions, for which the observed fluctuations within the NMR ensemble are not meaningful,
and which are usually absent from structures determined by X-ray crystallography. 

The structures in each NMR ensemble were superposed,
so as to ensure the absence of any rigid-body component to the observed displacements of the atomic coordinates.
Pseudo B-factors, corresponding to thermal fluctuations due solely to internal motions, were then computed for each residue $i$: 

\begin{equation}
\mean{\left|\Delta \vec{r}_i\right|^2}_{\mathrm{int}} = \frac{1}{N_m} \sum^{N_m}_{m=1} \left|\vec{r}^{\:\mathrm{int}}_{mi} - \mean{\vec{r}^{\:\mathrm{int}}_i}\right|^2
\label{eq:NMRint}
\end{equation} 

\noindent
where $N_m$ is the number of structural models in the ensemble, $\vec{r}^{\:\mathrm{int}}_{mi}$ is
the position of the C$_{\alpha}$ atom of residue $i$ in model $m$ of the superposed ensemble,
and the averages are taken over all models of the ensemble.

\subsection*{Generation of simulated sets of B-factors}

On the basis of the NMR dataset, we generated $N_s=33$ sets of simulated data,
by adding rigid-body contributions of controlled amplitude to the thermal fluctuations.
For each set $s$, we applied the following procedure to all superposed NMR ensembles.
Each structure $m$ in the ensemble was subjected to a random translation and rotation,
and its coordinates were adapted in consequence:

\begin{equation}
\vec{r}^{\:\mathrm{s}}_{mi} = a_{\omega} \vec{\omega}_m\times \vec{r}^{\:\mathrm{int}}_{mi} + a_t \vec{t}_m
\label{eq:NMRint2}
\end{equation} 

\noindent
The orientations of the rotation and translation vectors $\vec{\omega}_m$ and $\vec{t}_m$ were 
drawn randomly from a uniform spherical distribution, and their amplitudes from a standard normal distribution.
The scalar parameters $a_{\omega}$ and $a_t$ give control over the relative importance of internal, rotational, and translational motions. 
The resulting structural ensemble can be considered as a series of snapshots of a molecule undergoing fluctuations due to both internal and rigid-body motions.
For each residue, the mean square fluctuations were then computed from this ensemble of snapshots,
ensuring that the contribution of rigid-body motions is affected by the same kind of noise as for the internal motions.

\begin{equation}
\mean{\left|\Delta \vec{r}_i\right|^2}_{\mathrm{s}}= \frac{1}{N_m} \sum^{N_m}_{m=1} \left|\vec{r}^{\:\mathrm{s}}_{mi} - \mean{\vec{r}^{\:\mathrm{s}}_i}\right|^2
\label{eq:NMRint3}
\end{equation} 

\noindent
The average fraction of motion due to internal degrees of freedom, for every protein in set $s$, is:
\begin{equation}
I_{s} = \frac{\sum^N_{i=1} \mean{\left|\Delta \vec{r}_i\right|^2}_{\mathrm{int}}}  {\sum^N_{i=1} \mean{\left|\Delta \vec{r}_i\right|^2}_{s}}
\label{eq:NMRint4}
\end{equation} 

\noindent
where $N$ is the number of atoms. The average fractions of motion due to translational ($T_{s}$) and rotational ($R_{s}$) degrees of freedom are defined similarly.
For each protein in each set, the parameters $a_{\omega}$ and $a_t$ were adjusted so as to reach pre-defined values of $I_{s}$, $T_{s}$, and $R_{s}$.
More precisely, simulated sets were build for 11 different $I_s$ values $\{0.1,0.16,0.2,0.24,0.3,0.4,0.5,0.6,0.7,0.8,0.9\}$,
with either $T_s=0$, $R_s=0$, or $T_s=R_s$.
The relative amplitude of added rigid-body motions, $(T_s+R_s)/I_s$, was thus varied from 11\% ($I_s=0.9$) to 900\% ($I_s=0.1$).

Note that the structural variability within a superposed NMR ensemble does not perfectly reflect 
the actual dynamical behaviour of the protein. Indeed, it may also be affected by the resolution of the NMR experiment,
and the way the structure-building software deals with missing or conflicting distance constraints.
Still, mean square fluctuations extracted from superposed NMR ensembles have been shown to correlate well with B-factors from X-ray experiments,
and with NMR measurements more directly related to protein dynamics \cite{Berjanskii2006}.
The structural variability within NMR ensembles has also been successfully used to investigate the behaviour of ENMs \cite{Lezon2010}, or to parametrize their force field \cite{Dehouck13}. 
In any case, for our purposes, it is not necessary to assume that the collection of structural models within an ensemble 
gives an accurate picture of the protein's dynamics.
We merely consider that each ensemble provides a reasonably realistic example of possible fluctuations due to internal motions,
captured with a certain level of noise.

\subsection*{Elastic Network Model}
In this work we adopted the torsional network model (TNM), an ENM in torsion angle space that preserves the bond lengths
and bond angles of the protein \cite{TNM}. All protein atoms were considered in the computation of the kinetic energy.
The native interactions were identified with pairs of heavy atoms at distance smaller than $C$, which was varied from $3.0$ to $5.5$\AA.
For every pair of residues, only the pair of atoms at smallest distance were regarded as native contacts and joined
with a spring with force constant $\kappa(r)=\kappa_0(r_0/r)^E$, where $r$ is the equilibrium distance between the two atoms,
$r_0=3.5$\AA  is a reference distance, $\kappa_0$ is the force constant obtained from the fit of B-factors
and $E$ is an exponent that we varied from 0 to 8. Finally, the force constant of torsion angles was fixed at $\kappa_\phi=\kappa_\psi=0.1$,
a value that we had previously tested as almost optimal.

\subsection*{Quantities used for assessment}
The relative error of the fit is given by:

\begin{equation}
E = \frac{\sum^N_{i=1} (y_i - \sum^{P}_{k=1} X_{ik} a_k)^2}  {\sigma^2_{y}}
\label{eq:errorFit}
\end{equation} 

\noindent
where $N$ is the number of atoms and $P$ the number of predictor variables ($P=11$ when rigid-body motions are accounted for).
The $X_k$ are the normalised and dimensionless predictor variables, $a_k$ the corresponding fitted parameters,
and $\sigma^2_{y}$ is the variance of the dependent variable $\mathbf{y}=\mean{\left|\Delta \vec{r}\right|^2}$.

In the simulated sets, the contributions of the internal degrees of freedom are known exactly.
A straightforward way to evaluate the quality of the different fitting procedures is thus to assess their ability to accurately
extract the atomic fluctuations due to internal motions, $\mathbf{y}^{\mathrm{int}}=\mean{\left|\Delta \vec{r}\right|^2}_{\mathrm{int}}$,
from datasets with varying amounts of added "noise" (i.e. rigid-body fluctuations).   
The relative error on internal motions is defined as:

\begin{equation}
E^{\mathrm{int}} = \frac{\sum^N_{i=1} (y^{\mathrm{int}}_i - X_{i,\mathrm{ENM}} a_{\mathrm{ENM}})^2}  {\sigma^2_{y^{\mathrm{int}}}}
\label{eq:errorInt}
\end{equation} 

\noindent
where the $a_{\mathrm{ENM}}$ parameter is obtained from the full fit of the atomic fluctuations $\mean{\left|\Delta \vec{r}\right|^2}_s$.
A related measurement is obtained by considering the force constant estimated
from the NMR set without rigid-body motions, $\kappa_0$, as a proxy of the real force constant.
Considering that the force constant $\kappa$ is a multiplicative factor in the ENM model, 
the $\kappa$ values derived from the fits on simulated sets with added rigid-body motions are compared to $\kappa_0$ as follows: 

\begin{equation}
E^{\kappa} = \left|\log(\kappa/\kappa_0)\right|
\label{eq:errorKappa}
\end{equation} 

\noindent
The error measures are computed for each protein independently, 
and the reported values of $E$ and $E^{\mathrm{int}}$ are averaged over all proteins in the considered dataset.
Since $E^{\kappa}$ is not defined when $\kappa<0$, the reported values are averaged over the subset of proteins
for which none of the fitting procedures generates a negative value of $\kappa$
(i.e. between 165 ($I=0.1$) and 176 ($I=1.0$) proteins, out of 183, for the different simulated sets).

In the simulated datasets, the fractions of motion
due to internal ($I$), translational ($T$), and rotational ($R$) degrees of freedom were adjusted to specific predefined values.
The corresponding contributions of the three types of degrees of freedom can be estimated from the different fits.
In particular, the fitted fraction of motion due to internal degrees of freedom, in protein $p$, is:

\begin{equation}
I_{p} = \frac{1}{N} \sum^{N}_{i=1} \frac{X_{pi,\mathrm{ENM}} a_{p,\mathrm{ENM}}}{\sum^{P}_{k=1} X_{pik} a_{pk}}
\label{eq:fittedI}
\end{equation} 

\noindent
For each fit, we report the RMSE between the fitted and real fractions of motion due to internal degrees of freedom:

\begin{equation}
\mathrm{RMSE}(I_p)= \sqrt{\frac{1}{N_p} \sum_{p=1}^{N_p} (I_p - I)^2}
\label{eq:RMSEI}
\end{equation} 

\noindent
where $N_p$ is the number of proteins in the dataset.
The corresponding measures for the translational and rotational degrees of freedom, RMSE($T_p$) and RMSE($R_p$), respectively, are defined similarly.


\begin{thebibliography}{9}

\bibitem{Kern2005}
Eisenmesser EZ, Millet O, Labeikovsky W, Korzhnev DM, Wolf-Watz M, Bosco DA, Skalicky JJ, Kay LE, Kern D. (2005)
Intrinsic dynamics of an enzyme underlies catalysis.
Nature 438:117-21.

\bibitem{Goodey_Benkovic_2008}
Goodey NM, Benkovic SJ. (2008)
Allosteric regulation and catalysis emerge via a common route.
Nat Chem Biol. 4:474-82.

\bibitem{allostery}
del Sol A, Tsai CJ, Ma B, Nussinov R. (2009)
The origin of allosteric functional modulation: multiple pre-existing pathways.
Structure 17:1042-50.

\bibitem{conf_selection}
Boehr DD, Nussinov R, Wright PE. (2009)
The role of dynamic conformational ensembles in biomolecular recognition.
Nat Chem Biol. 5:789–96.

\bibitem{Kern2010}
Villali J, Kern D. (2010) Choreographing an enzyme's dance.
Curr Opin Chem Biol. 14:636-43.

\bibitem{Tirion96}
Tirion MM. (1996)
Large Amplitude Elastic Motions in Proteins from a Single-Parameter, Atomic Analysis.
Phys Rev Lett. 77:1905-1908.

\bibitem{Atilgan01}
Atilgan AR, Durell SR, Jernigan RL, Demirel MC, Keskin O, Bahar I. (2001)
Anisotropy of fluctuation dynamics of proteins with an elastic network model.
Biophys J. 80: 505–515.

\bibitem{Bahar_review_05}
Bahar I, Rader AJ. (2005)
Coarse-grained normal mode analysis in structural biology.
Curr Opin Struct Biol. 15:586-92.

\bibitem{Tama2001}
Tama F, Sanejouand YH. (2001)
Conformational change of proteins arising from normal mode calculations.
Protein Eng. 14:1-6.

\bibitem{binding}
Meireles L, Gur M, Bakan A, Bahar I. (2011)
Pre-existing soft modes of motion uniquely defined by native contact topology facilitate ligand binding to proteins.
Protein Sci. 20:1645-58.

\bibitem{Bastolla2013}
Dos Santos HG, Klett J, M\'endez R, Bastolla U. (2013)
Characterizing conformation changes in proteins through the torsional elastic response.
Biochem Biophys Acta 1834:836-46.

\bibitem{Orozco07}
Rueda M, Chac\'on P, Orozco M. (2007)
Thorough validation of protein normal mode analysis: a comparative study with essential dynamics.
Structure 15:565-75.

\bibitem{CHARMMvsENM}
Kondrashov DA, Van Wynsberghe AW, Bannen RM, Cui Q, Phillips GN Jr. (2007)
Protein structural variation in computational models and crystallographic data.
Structure 15:169-77.

\bibitem{Hinsen00}
Hinsen K, Petrescu AJ, Dellerue S, Bellisent-Funel MC, Kneller GR. (2000)
Harmonicity in slow protein dynamics.
Chem Phys. 261: 25–37.

\bibitem{Moritsugu07}
Moritsugu K, Smith JC. (2007)
Coarse-grained biomolecular simulation with REACH: realistic extension algorithm via covariance Hessian.
Biophys J. 93: 3460–3469.

\bibitem{Yang09}
Yang L, Song G, Jernigan RL. (2009)
Protein elastic network models and the ranges of cooperativity.
Proc Natl Acad Sci (USA) 106: 12347–12352.

\bibitem{Dehouck13}
Dehouck Y, Mikhailov AS. (2013)
Effective harmonic potentials: insights into the internal cooperativity and sequence-specificity of protein dynamics.
PLoS Comput Biol. 9:e1003209.

\bibitem{Hamacher06}
Hamacher K, McCammon J. (2006)
Computing the amino acid specificity of fluctuations in biomolecular systems.
J Chem Theory Comput. 2: 873–878.

\bibitem{Gerek09}
Gerek ZN, Keskin O, Ozkan SB. (2009)
Identification of specificity and promiscuity of pdz domain interactions through their dynamic behavior.
Proteins 77: 796–811.

\bibitem{TNM}
Mendez R, Bastolla U. (2011)
Torsional network model: normal modes in torsion angle space better correlate with conformation changes in proteins.
Phys Rev Lett. 2010 104:228103.

\bibitem{Kuriyan1991}
Kuriyan J, Weis WI. (1991)
Rigid protein motion as a model for crystallographic temperature factors.
Proc Natl Acad Sci (USA) 88:2773-2777.

\bibitem{Hinsen2008}
Hinsen K (2008)
Structural flexibility in proteins: impact of the crystal environment.
Bioinformatics 24:521-528.

\bibitem{Riccardi2009}
Riccardi D, Cui Q, Phillips GN Jr. (2009)
Application of elastic network models to proteins in the crystalline state.
Biophys J. 96:464–475.

\bibitem{Soheilifard2008}
Soheilifard R, Makarov  DE, Rodin GJ. (2008)
Critical evaluation of simple network models of protein dynamics and their comparison with crystallographic B-factors.
Phys Biol. 5:026008.

\bibitem{Hafner2010}
Hafner J, Zheng W. (2010)
Optimal modeling of atomic fluctuations in protein crystal structures for weak crystal contact interactions.
J Chem Phys. 132:014111.

\bibitem{Lezon2012}
Lezon TR. (2012)
The effects of rigid motions on elastic network model force constants.
Proteins 80:1133-1142.

\bibitem{TLS}
Schomaker V, Trueblood KN. (1968)
On the rigid-body motion of molecules in crystals.
Acta Crystallogr. B. 24:63–76.

\bibitem{ridge}
Hoerl AE, Kennard RW. (1970)
Ridge regression: biased estimation for nonorthogonal problems.
Technometrics 12:55-67.

\bibitem{GCV}
Golub GH, Heath M, Wahba G. (1979)
Generalized cross-validation as a method for choosing a good ridge parameter.
Technometrics 21:215-223.

\bibitem{RR}
Mallows C. (1973)
Some comments on $C_p$.
Technometrics 15:661-675.

\bibitem{L-curve}
Hansen PC. (1992)
Analysis of discrete ill-posed problems by means of the L-curve.
SIAM Rev. 34:561-580.

\bibitem{Riccardi2010}
Riccardi D, Cui Q, Philips GN Jr. (2010)
Evaluating elastic network models of crystalline biological molecules with temperature factors, correlated motions, and diffuse x-ray scattering.
Biophys J. 99:2616-25.

\bibitem{pisa}
Krissinel E, Henrick K. (2007)
Inference of macromolecular assemblies from crystalline state.
J Mol Biol. 372:774-797.

\bibitem{MolProbity}
Davis IW, Leaver-Fay A, Chen VB, Block JN, Kapral GJ, Wang X, Murray LW, Arendall WBIII, Snoeyink J, Richardson JS, Richardson DC. (2007)
MolProbity: all-atom contacts and structure validation for proteins and nucleic acids. 
Nucleic Acids Res. 35: W375–W383.

\bibitem{Berjanskii2006}
Berjanskii M, Wishart DS. (2006)
NMR: prediction of protein flexibility.
Nat Protoc. 1:683-8.

\bibitem{Lezon2010}
Lezon TR, Bahar I. (2010)
Using entropy maximization to understand the determinants of structural dynamics beyond native contact topology.
PLoS Comput Biol. 6:e1000816.

\bibitem{Kuzmanic2014}
Kuzmanic A, Pannu NS, Zagrovic B. (2014)
X-ray refinement significantly underestimates the level of microscopic heterogeneity in biomolecular crystals.
Nat Commun. 5:3220.

\bibitem{Fuglebakk2012}
Fuglebakk E, Echave J, Reuter N. (2012)
Measuring and comparing structural fluctuation patterns in large protein datasets.
Bioinformatics 28:2431–2440.

\bibitem{Fuglebakk2013}
Fuglebakk E, Reuter N, Hinsen K. (2013)
Evaluation of protein elastic network models based on an analysis of collective motions.
J Chem Theor Comput. 9:5618-5628.

\bibitem{Halle2002}
Halle B. (2002)
Flexibility and packing in proteins.
Proc Natl Acad Sci (USA).
99:1274-9.

\bibitem{Li2009}
Li DW, Br\"uschweiler R. (2009)
All-atom contact model for understanding protein dynamics from crystallographic B-factors.
Biophys J. 96:3074-81.

\end{thebibliography}
\end{document}